\documentclass[a4paper,12pt]{article}
\pdfoutput=1
\def\parn              {  \par\noindent }

\def\parmedskipn        {  \par\medskip\noindent  }

\def\ep{\epsilon}

 \def\calH{{\cal H}} 
  
  \def\calO{{\cal O}}

\def\Atil{\widetilde{A}}
\def\Btil{\tilde{B}}
\def\Ctil{\widetilde{C}}
\def\Dtil{\widetilde{D}}

\def\Ktil{{\tilde{K}}}
\def\Ltil{\widetilde{L}}

\def\del        {  \partial }

\def\rootof#1   {  \left( #1 \right)^{1/2}  } 

\def\Tr         { {\rm Tr}\,  }
\def\abs#1      {  \vert #1 \vert  }
\def\ie         {{\it i.e.}\,\,}
\def\evalat#1   {  \left\vert_{#1} \right. }

\def\comma          {\, ,}
\def\period         {\, .}
\def\lsim      {\lower .65ex \hbox{\ $\stackrel{<}{\sim}$\ } }
\def\gsim      {\lower .65ex \hbox{\ $\stackrel{>}{\sim}$\ } }
\def\det       {{\rm det}\, }

\def\bra#1{{\langle #1 | } }
\def\ket#1{{| #1 \rangle } }
\def\braket#1#2{\langle #1 | #2 \rangle}

\def\matel#1#2#3  {{\langle #1 | #2 | #3 \rangle } }

\def\lrvec#1    {\hbox{$\stackrel{\leftrightarrow}{#1}$}}
\def\lvec#1     {\hbox{$\stackrel{\leftarrow}{#1}$}}

\def\vecii#1#2      {  \left(\begin{array}{c}#1\\#2\end{array}\right)  }
\def\veciii#1#2#3   {  \left(\begin{array}{c}#1\\#2\\#3\end{array}
                     \right)  }
\def\veciv#1#2#3#4  {  \left(\begin{array}{c}#1\\#2\\#3\\#4
                                 \end{array}\right)  }
\def\vecfv#1#2#3#4#5 {  \left(\begin{array}{c}#1\\#2\\#3\\#4\\#5
                                 \end{array}\right)  }

\def\matrixii#1#2#3#4            {  \left(\begin{array}{cc}#1&#2\\#3&#4
                                       \end{array}\right) }
\def\matrixiii#1#2#3#4#5#6#7#8#9 {  \left(\begin{array}{ccc}#1&#2&#3\\
                                     #4&#5&#6\\#7&#8&#9\end{array}
                               \right)  }
\def\mativ#1#2#3#4               {  \left(\begin{array}{cccc}
                                       #1\\#2\\#3\\#4\end{array}\right) }
\def\matv#1#2#3#4#5              {  \left(\begin{array}{ccccc}
                                     #1\\#2\\#3\\#4\\#5\end{array}
                              \right)  }
\def\eqabegin         {  \begin{eqnarray}  }
\def\eqaend           {  \end{eqnarray}  }
\def\nn               {  \nonumber  }
\def\bracetwo#1#2     {  \left\{ \begin{array}{l} #1 \\ #2 \end{array}
                         \right.  }
\def\bracetwocases#1#2#3#4  {   \left\{ \begin{array}{ll} #1 &
                                 \qquad #2 \\
                                 #3 & \qquad #4 \end{array} \right.  }
\def\bracebegin#1     {  \left\{ \begin{array}{#1}   }
\def\braceend         {  \end{array}\right.   }
\def\shead#1   { \parmedskipn {\bfall $\Box$\ #1}: \parmedskipn }
\def\head#1    { \parmedskipn {\bfall $\Box$\ #1}: \qquad }
\def\lhead#1   { \parmedskipn {\large\bfall $\Box$\ #1} \parmedskipn }
\def\Lhead#1   { \parmedskipn {\Large\bfall $\Box$\ #1} \parmedskipn }
\def\boxit#1#2      {  \vbox{\hrule\hbox{ \hskip -4.1pt \vrule\kern3pt 
                     \vbox
                    {  \hsize #1 \strut\kern3pt #2 \kern3pt\strut  }
                       \kern3pt  \vrule} \hrule  } }
\def\centerbox#1#2  {  \mbox{  }\par\bigskip  \hfil \boxit{#1}{#2} \hfil
                       \par\bigskip\noindent }
\def\rightbox#1#2   {  \hfill\boxit{#1}{#2}  }
\def\leftbox#1#2    {  \boxit{#1}{#2}  }
\def\fullbox#1      {  \boxit{\textwidth}{#1}  }

\newcommand{\nullify}[1]{}

\def\mpg#1#2{\begin{minipage}[t]{#1} #2  \end{minipage} }

\def\bfall{\boldmath\bf }

\def\epsfig#1#2#3{
{\lower #3 \hbox{
 \mpg{#1}{\begin{center} \includegraphics[width=#1,clip]{#2.eps} \\
 Fig. #2\end{center} }}}}

\def\papertitlepage{\baselineskip 3.5ex \thispagestyle{empty}}
\def\Title#1{\baselineskip 1cm \vspace{1.5cm}\begin{center}
 {\Large\bf #1} \end{center} 
\vspace{0.5cm}}
\def\Authors#1{\begin{center} {\it #1} \end{center}}
\def\Abstract{\vspace{1.0cm}\begin{center} {\large\bf Abstract} 
           \end{center} \par\bigskip}
\def\Komabanumber#1#2#3{\hfill \begin{minipage}{4.2cm} UT-Komaba #1
              \parn #2 
              \parn #3 \end{minipage}}
\renewcommand{\thefootnote}{\fnsymbol{footnote}}

\renewenvironment{thebibliography}{\pagebreak[3]\par\vspace{0.6em}
\begin{flushleft}{\large \bf References}\end{flushleft}
\vspace{-1.0em}

\begin{enumerate}\if@twocolumn\baselineskip=0.6em\itemsep -0.2em
\else\itemsep -0.2em\fi\labelsep 0.1em}{\end{enumerate} }
\def\bfall{\boldmath\bf}
\usepackage{ifpdf}
\ifpdf
\usepackage{graphicx} 
\usepackage{epsfig}
\usepackage[setpagesize=false,pagebackref=false, linktocpage, bookmarksopen=true, colorlinks=true, linkcolor=darkblue,citecolor=darkblue,urlcolor=darkblue]{hyperref}
\usepackage{hyperref}
\newcommand{\arXiv}[2]{\href{http://arxiv.org/abs/#1}{{\tt arXiv:#2}}}
\newcommand{\hep}[2]{\href{http://arxiv.org/abs/#1}{{\tt #2}}}

\else
\usepackage{graphicx}
\newcommand{\arXiv}[2]{{\tt arXiv:#2}}
\newcommand{\hep}[2]{{\tt #2}}

\fi
\usepackage{latexsym}
\usepackage{amsmath}
\usepackage{amssymb}
\usepackage[usenames]{color}
\usepackage{cite}
\usepackage{fancybox}

\definecolor{darkgreen}{rgb}{0,0.5,0}
\definecolor{darkblue}{rgb}{0.1,0.1,0.7}
\definecolor{darkred}{rgb}{0.6,0,0.3}
\definecolor{MyRed}{cmyk}{0,1,1,0.15}
\definecolor{MyBlue}{cmyk}{1,1,0,0.25}

\def\fn#1{\footnote{#1}}
\def\eqref#1{(\ref{#1})}

\def\bsv{\boldsymbol{v}}
\def\bsu{\boldsymbol{u}}

\def\bsth{\boldsymbol{\theta}}

\def\Qu{Q_{\bsu}}
\def\Qv{Q_{\bsv}}

\def\Qt{Q_{\bsth}}
\def\beq#1{\begin{align}#1\end{align}}
\def\pmatrix#1#2{\left( 
\begin{array}{#1}
#2\end{array} 
\right)}
\def\id{\text{\boldmath $1$}}
\def\upvacket{\ket{\uparrow^L}}
\def\upvacbra{\bra{\uparrow^L}}
\def\xhat{\hat{x}}
\def\yhat{\hat{y}}
\def\upket{\ket{\uparrow^L}}
\def\upbra{\bra{\uparrow^L}}

\begin{document}
\papertitlepage
\vspace*{0cm}
\Komabanumber{13-6}{April, 2013}{}
\Title{A new integral representation for the scalar products \\
of Bethe states for the XXX spin chain } 
\Authors{{\sc Yoichi Kazama\footnote[2]{\href{mailto:kazama@hep1.c.u-tokyo.ac.jp}{\texttt{kazama@hep1.c.u-tokyo.ac.jp}}}, Shota Komatsu\footnote[3]{\href{mailto:skomatsu@hep1.c.u-tokyo.ac.jp}{\texttt{skomatsu@hep1.c.u-tokyo.ac.jp}}} and Takuya Nishimura\footnote[4]{\href{mailto:tnishimura@hep1.c.u-tokyo.ac.jp}{\texttt{tnishimura@hep1.c.u-tokyo.ac.jp}}}
\\ }
\vskip 2ex
 Institute of Physics, University of Tokyo, \\
 Komaba, Meguro-ku, Tokyo 153-8902 Japan \\
  }
\baselineskip .7cm

\numberwithin{equation}{section}
\numberwithin{figure}{section}
\numberwithin{table}{section}
\parskip=0.9ex

\Abstract
Based on the method of separation of variables due to Sklyanin, 
we construct a new  integral representation for the scalar  products of 
 the Bethe states for the SU(2) XXX spin $1/2$ chain obeying  the periodic
 boundary condition.  Due to the compactness of the symmetry group, 
 a  twist matrix  must be introduced at  the boundary 
in order to  extract the separated variables properly. Then by deriving the 
integration measure and the spectrum of the separated variables, 
we  express the inner product of an on-shell and an off-shell 
Bethe states in terms of a  multiple contour integral involving a product 
 of Baxter wave functions.  Its form is reminiscent of the integral 
 over the eigenvalues of a matrix model and is expected to be useful in 
 studying the semi-classical limit of the product. 

\newpage
\baselineskip 3.5ex
\renewcommand{\thefootnote}{\arabic{footnote}}
\setcounter{footnote}{0}
\thispagestyle{empty}
\enlargethispage{2\baselineskip}
\hrule
\tableofcontents
\vspace{11pt}
\hrule
\section{Introduction }
The quantum XXX Heisenberg spin chain is undoubtedly one of the 
 most celebrated quantum integrable systems of fundamental interest. 
Besides being the prototypical model of the magnetic substance  in the condensed 
 matter physics, it has been under intense study as well from the standpoint of 
 mathematical physics of exactly solvable models \cite{book}.  
As such,  innumerable studies have been made in the past and the basic properties  of the model are  thought to be well understood. 

However, about ten years ago  this system started to receive a renewed interest 
from quite a different perspective as it made its  appearance in  an unexpected 
way: It was recognized as the precise  mathematical structure governing 
the scaling properties  of the gauge-invariant single-trace composite operators in certain sectors of  the $\mathcal{N}=4$ SU(N) super Yang-Mills theory at the one-loop level
\cite{MZ}.  In the simplest 
 situation, where the composite operators are made out of two kinds of  adjoint-valued  complex scalars $Z$ and $X$, $Z$ ($X$)  can be identified with the spin up (down)  state of an individual spin forming an SU(2) spin chain and the dilatation 
 operator acting on such composite operators takes exactly the form 
 of the well-known XXX spin chain Hamiltonian. This identification allows  one to 
 compute the eigenstates and the anomalous dimensions of the composite 
 operators using the techniques of the integrable models, such as the Bethe 
 ansatz \cite{review}. This technique is particularly useful when the number of magnon excitations as well as
 the number of spins  become very large. In such a semi-classical 
 limit, one can recognize the integrable structure quite similar to that 
of a classical string in a curved spacetime containing an AdS subspace and, 
as far as the spectrum of the excitation is concerned,  this 
provides a structural parallel strongly suggesting  the AdS/CFT correspondence
\cite{review}. 

More recently, further integrable properties of the XXX spin chain have been exploited  beyond the spectral level to compute the correlation functions of the composite operators of the super Yang-Mills theory. Namely,  in a series of papers \cite{EGSV1, EGSV2, EGSV3, GV, EGSV4}, 
the method of computing the three point functions $\langle \calO_1(x_1) \calO_2(x_2) 
\calO_3(x_3) \rangle$ has been developed, 
built on earlier works \cite{OT, RV, ADGN},  in such a way that 
the three point function can be expressed in terms of the scalar products 
 of the  Bethe states of the spin chain Hamiltonian.  For relevant 
 configurations, such scalar products can be expressed in terms of the so-called 
 Slavnov determinants \cite{Slavnov, Foda}, which can be further simplified for special cases 
 where BPS operators are involved.  
 
Now in order to study the structural similarity with the three point 
 functions in the strong coupling regime represented by  semi-classical string 
in  AdS spacetime\footnote{The one corresponding to  the 
 SU(2) sector of the present interest is the string in $AdS_2 \times S^3$. 
The contribution from the $AdS_2$ part has been obtained in \cite{JW}. The contribution from the $S^3$ part, which is more involved, 
 will be presented in a forthcoming paper \cite{KK3}. }, it is of  importance  to consider the semi-classical limit where the number of the magnon excitations as well as spins become very large. 
For the case where one of the operators is non-BPS and the others are BPS, such a limit has been obtained in a remarkably compact form \cite{EGSV3}. More recently, the semi-classical limit of the fully non-BPS 
three point function has been worked out by Kostov \cite{Kostov, Kostov2}. This was achieved again 
 starting from the expression in terms of the Slavnov's determinant 
 formula in the framework of the algebraic Bethe ansatz, which was 
 quite ingenious. 

 One of the strong motivations for the study performed in 
 this paper is to construct  a representation of the scalar products with  which 
 the semi-classical limit may be understood more physically. 
We believe that such an understanding should be important in order to seek  the holographic string picture in the super Yang-Mills theory. 

As has been already indicated, studies of the XXX spin $1/2$ chain 
in the past have been performed 
predominantly in the framework of the algebraic Bethe ansatz. 
The main ingredients of this method are the elements of the monodromy 
 matrix  $\Omega(u) =\matrixii{A(u)}{B(u)}{C(u)}{D(u)} $, 
which are themselves operators acting on the spin chain Hilbert space, 
 with  up and down spins at, say,   $L$ sites. The operators $A(u), \ldots ,D(u)$ 
 satisfy the Yang-Baxter exchange algebra\footnote{We shall only recall  a necessary portion of this algebra later when we need them.}  and from this one 
 sees that $B(u)$ and $C(u)$ can be regarded as the ``creation" and the ``annihilation" operators  respectively,   relative  to the ``vacuum" $\upvacket$ with 
 all the spins up. Thus one can construct the basic  state with $M$ magnon 
 excitations in the form  $\prod_{i=1}^M B(u_i) \upvacket$ and 
 its conjugate $\upvacbra \prod_{i=1}^M C(u_i)$. 
Such a state will be referred to as a  Bethe state. It is well-known that a Bethe state becomes an eigenstate of the mutually commuting conserved charges, among which is 
 the Heisenberg Hamiltonian of the spin chain,  when  
 the rapidities $u_i$ of the magnons  satisfy the Bethe equations, \ie when they are 
``on-shell".  Of fundamental importance  in this frame work is the scalar product 
$\upvacbra  \prod_{i=1}^M C(v_i)\prod_{i=1}^M B(u_i) \upvacket$, which 
 can be computed using the Yang-Baxter algebra. When one of the set of  rapidities, for instance $\{v_i\}$,  are on-shell,  the scalar product can be 
 simplified  enormously and  expressed as a determinant. This is the 
 celebrated Slavnov determinant and practically all  the calculations 
 involving the scalar product have been done starting from this expression. 

In this article, however, we shall take a different route for the calculation of 
 this scalar product and naturally obtain a different new representation. 
This alternative is the method of separation of variables (SoV), which was 
 advanced 
substantially by  Sklyanin \cite{Sklyanin}.
 The concept of SoV represents  the most primitive and 
 fundamental form of  integrability, where one reduces the interacting 
many-body system to mutually decoupled set  of  
 dynamical systems, each with  a single degree of freedom. Of course the 
 highly non-trivial question is how to actually construct  
 such  separated variables $\{x_k\}$ and the 
corresponding canonically conjugate momenta $\{p_k\}$. For the integrable 
 systems which admit the formulation with Lax operators, Sklyanin proposed 
 a powerful concrete recipe for the construction.  Relegating  more detailed 
 description to section \ref{sec-3}, the prescription applied to the case of XXX spin 1/2 
 chain says that the solutions  $x_k$ of the operator equation $B(x_k)=0$ 
provide the separated coordinates, while their  conjugate momenta $p_k$ 
 are given essentially by $D(x_k)$.  One can indeed check that they satisfy 
 (with appropriate ordering in the quantum case) the canonical Poisson-Lie 
 commutation relations. Therefore if one can diagonalize and factorize $B(u)$
 as $B(u) \propto \prod_{k=1}^L (u-x_k)$, with precisely $L$ zeros, $x_k$
  provide a complete basis of separated coordinates. Once this is achieved, 
 one can figure out the measure factor $\mu(x_1,\ldots,x_L)$ and 
 compute the scalar products between various states in  the $x$-representation. 

Indeed such a method has been applied successfully to some cases where the conventional algebraic Bethe ansatz  is not readily applicable. 
One example is the non-compact SL(2) spin chain in the unitary representation,  studied in \cite{sl(2)}.  One gratifying  feature of this case is that 
in such a  unitary representation  the hermitian conjugate of $B(u)$ 
operator is basically itself and hence can be easily diagonalized. The integration 
 measure is found and the scalar product is thus defined in the SoV 
framework. 
 Another system for which the SoV analysis  has been  performed
is  the SU(2) spin chain with anti-periodic boundary condition \cite{Niccoli1, Niccoli3}. 
In this case, due to the insertion  of the twisting matrix  $K=\matrixii{0}{1}{1}{0} $ which flips the 
 spin at the boundary, the operator which should be diagonalized to yield 
 separated variables changes from $B(u)$ to  $D(u)$. 
  Since this operator is hermitian and naturally diagonalizable 
 the subsequent analysis \`a la Sklyanin is straightforward.

Now for the more fundamental case of the SU(2) spin chain
 with the periodic boundary condition,  there are two apparent obstacles 
 in computing the scalar products using  the Sklyanin's procedure. 
The first problem is that  because the hermitian conjugate of $B(u)$ is 
 $C(u)$,  the basis in which $B(u)$ is diagonal is different from 
 the one in which $C(u)$ is diagonal. Hence the scalar product of our interest 
$\langle \bsv \ket{\bsu} =\upvacbra  \prod_{i=1}^M C(v_i)\prod_{i=1}^M B(u_i) \upvacket$ 
cannot be easily computed in $B(u)$-diagonal basis. 
The second problem is that $B(u)$ operator as it stands is actually 
not a good operator in the SoV framework, 
since the coefficient of the highest power $u^L$ in the expansion of 
 $B(u)$ is proportional to $S^-=S^x-iS^y$  belonging to the 
 global SU(2), which is obviously not diagonalizable. 
It is perhaps for these reasons that this important basic model has not been 
 treated in the SoV basis so~far. 

We will  solve these problems in the following manner. As for the first problem, 
 since we are interested in the case where $\{v_i\}$ are on-shell, we may 
 use the trick due to Kostov and Matsuo \cite{KM} to rewrite the scalar product 
 into the form 
 $\sim \bra{\downarrow^L} (S_-)^{L-2M} \prod_{i=1}^M B(v_i)  B(u_i) \ket{\uparrow^L}$, where only the $B(u)$ operators appear. Then, 
the second problem can be solved by introducing  a 
boundary condition changing twisting matrix $K_\ep = \matrixii{1}{\ep}{-\ep}{1} $ so that the modified (regularized) operator $B_\ep(u)$ is diagonalizable. 
As we shall describe in detail in section \ref{sec-4}, we can compute the integration measure as well as the wave functions corresponding to the general Bethe states 
$ \prod_{i=1}^N B(w_i)\upvacket$  in the SoV basis. When put together to 
 form the scalar product $\ep$-dependence in various quantities cancel precisely. 
This is as it should be since the original 
scalar product $\langle {\bsv}\ket{\bsu}$ is completely well-defined and finite. 
The original representation we obtain this way consists of contour integrals over $x_k$ which surround certain poles of the integrand that depend on the index $k$. 
This can be recast  into a more convenient form 
 where the integration contours for all the $x_k$'s will encircle all the simple poles of the integrand\fn{Multiple integral formulas for the scalar product exist in 
 the literature\cite{GGS, Galleas}. Our formula differs from them in form as well
as in the context in which it is derived.}. This expression resembles the integral over the eigenvalues of a matrix model and is expected to be useful in studying the semi-classical limit of the scalar product. In the Appendix \ref{ap-b}, we shall give a direct  independent  proof  that 
our  integral representation for the $2M=L$ case is equivalent to the  Izergin's determinant formula for the so-called domain wall partition function of  the 6-vertex model\footnote{Although we will not emphasize it in this article, 
regarding the XXX spin $1/2$ chain from the point of view of the 6-vertex model provides certain useful insights and there have been many interesting works on this topic \cite{Foda,FW1, FW2}. }.  Using the Kostov-Matsuo trick and taking appropriate limits for the rapidities on both sides, we obtain  an alternative proof that our integral representation reproduces 
 the Slavnov's determinant formula for the scalar product of interest.

The rest of this article is organized as follows. We begin  in section \ref{sec-2} by 
giving a brief review of the framework of the algebraic Bethe ansatz 
where the scalar products of interest are defined and describe 
 several different forms of the  determinant formulas for them 
and related quantities. In section \ref{sec-3} the essence of the Sklyanin's method of  separation of  variables  for classical and quantum integrable models will be 
summarized. With these preparations, we will derive in section \ref{sec-4} a new integral
 representation  for the scalar product  between an on-shell and an off-shell Bethe states for the XXX spin $1/2$ chain in a  separated variable basis. 
The remaining problems to be pursued, in particular  that of  deriving the semi-classical limit from our integral representation,  will be briefly 
 discussed in section \ref{sec-5}. Two appendices will be provided to give some technical 
details of the derivation. 
\section{Algebraic Bethe ansatz and determinant formulas \label{sec-2}}
As a preliminary, we shall give a brief review of the algebraic Bethe ansatz for the 
 XXX spin 1/2 chain and summarize the various known forms of the 
 determinant type formulas for the scalar products between 
 the Bethe states, the quantity of our prime interest. This will at the same time 
serve to  fix our notations. 
\subsection{Algebraic Bethe ansatz}
The basic ingredient  in the  framework of the algebraic Bethe ansatz is the 
so-called Lax operator acting on the  product of the spin-chain 
 Hilbert space $\calH$ and an auxiliary vector space. In the case of  the 
 XXX spin $1/2$ chain with $L$ sites,  $\calH$ is 
  the tensor product of  $L$ copies of a two-dimensional vector space, 
consisting of the up-spin state $|\uparrow\,  \rangle$ and the down-spin state $|\downarrow\,  \rangle$ at each site, and the auxiliary space has the structure of $\mathbb{C}^2$. The Lax operator $L_n(u)$ acting on the $n$-th  site 
 is then given by 
\begin{align} L_n(u) \equiv  u{\bf 1} +i\sum_{k=x,y,z} S_n^k \sigma^k =
\left( 
\begin{array}{cc}
u+iS_n^z & iS_n^-  \\
iS_n^+ & u-iS_n^z  
\end{array} 
\right)  \comma 
\end{align}
where $S^k_n$ are the local spin operators\footnote{We define $S^\pm_n$ 
 as $S^\pm_n \equiv S^x_n \pm i S^y_n$.}  and $u$ is the complex 
spectral parameter. We will impose the periodic boundary condition so that 
 $S^k_{n+L} = S^k_n$. Going around the spin chain,  we  define 
 the monodromy matrix $\Omega(u)$ as 
\begin{align}
\Omega (u)& \equiv  L_1(u-\theta_1 )\cdots L_L(u-\theta_L ) 
\equiv \left( 
\begin{array}{cc}
A(u) & B(u)  \\
C(u) & D(u) \\
\end{array}  
\right) \\
&= u^L {\bf 1} +iu^{L-1}\left(\sum_{k=x,y,z} S^k \sigma^k 
+ i \sum_{j=1}^L \theta_j \right)
+\mathcal{O}(u^{L-2})  \period 
 \label{eq:mon}
\end{align}
Here $S^k=\sum_{n}S_n^k$ are the total spin operators and 
 we have introduced the inhomogeneity parameters ${\boldsymbol \theta}=\{ \theta_1,\ldots ,\theta_L \}$ at each site, 
which preserve the integrability. They are necessary for avoiding certain 
 degeneracies in the intermediate steps and are also useful for other 
 purposes\footnote{Although the physical meaning of the inhomogeneity 
 parameters in the context of the three point functions has not been 
fully clarified, they are useful in generating loop corrections from the 
 tree-level contributions \cite{GV, EGSV4}.}.

Although the actions of the operators $A(u) \ldots D(u)$ on $\calH$ 
are in general quite complicated and non-local, they are known to satisfy 
rather simple exchange relations, which we call Yang-Baxter algebra \cite{book}. 
In particular its structure reveals that $B(u)$ and $C(u)$ can be interpreted 
 as a ``creation" and an ``annihilation" operator respectively with respect 
to the pseudovacuum  $| \uparrow^L \rangle \equiv \underbrace{ | \uparrow \rangle \otimes \cdots \otimes | \uparrow \rangle }_{L}$, in which  all the spins 
 are up. 
This allows one to construct the Hilbert space $\calH$ as the 
 Fock space spanned by the $M$-magnon states
 of the form $\ket{\boldsymbol{u}} =B(u_1) B(u_2) \cdots  B(u_M) \upket$, 
while $C(v) \upket =0$. $u_i$'s are the rapidities of the magnons, which are 
 related to the momenta by $p = \log {u+i/2 \over u-i/2}$.  
Similarly, the bra states are generated by the operator $C(v)$'s as 
$\bra{\bsv} = \upbra C(v_1) C(v_2) \cdots C(v_M)$, built upon 
 the dual pseudovacuum $\upbra$ satisfying $\upbra B(u) =0$ and 
 $\upbra \uparrow^L \rangle =1$. These Fock states will be referred to as generic Bethe states. 

Of particular importance is the transfer matrix given by 
$T(u) \equiv \Tr \Omega(u) = A(u) +D(u)$, 
 which upon expanded in powers of $u$ generates all the 
mutually commuting  conserved quantities, including the Hamiltonian of the spin chain. The (dual) pseudovacuum is known to be the eigenstate of $T(u)$
 in the manner
\begin{align}
 A(u)| \uparrow^L \rangle &=\Qt^+(u)| \uparrow^L \rangle  \comma 
\qquad  D(u)| \uparrow^L \rangle =\Qt^-(u)| \uparrow^L \rangle \comma  \\
\langle \uparrow^L |A(u) &=\langle \uparrow^L |\Qt^+(u)\comma \qquad  \langle \uparrow^L |D(u) =\langle \uparrow^L |\Qt^-(u) \comma 
\end{align}
where  $\Qt$ functions are defined as\footnote{As in these definitions, each $+$ ( respectively  $-$) superscript on a function 
signifies that its argument is shifted by $+{i \over 2}$ (respectively  $-{i \over 2}$). According to this convention,   $\Qt^{++}(u)$ means $\Qt(u+i)$, etc.
When $\theta_k=0$, the functions $\Qt^\pm(u)$ 
 are often referred to as $a(u)$ (for $+$) and $d(u)$ (for $-$).  }
\begin{align}
\Qt(u) & \equiv \prod_{k=1}^L \left( u-\theta_k \right) \comma 
\qquad \Qt^\pm(u) \equiv \prod_{k=1}^L \left( u-\theta_k \pm {i \over 2}\right) \period\label{Qt}
\end{align}
Using this fact, the action of $T(u)$ on the generic Bethe state $\ket{\bsu} =\prod_{i=1}^M 
B(u_i)\upket$  can be computed by pushing $A(u)$ and $D(u)$ through $B(u_i)$'s using the exchange relations such as 
$(u-v)A(v)B(u)  = (u-v+i)B(u)A(v)-iB(v)A(u)$ and a similar one between $D(v)$ and $B(u)$.  One then finds 
 that $\ket{\bsu}$ becomes the eigenstate of $T(u)$ if and only if 
 the following sets of equations, called the Bethe ansatz equations, for the 
 rapidities are satisfied:
 \begin{align} \prod_{k=1}^L \left( \frac{u_j-\theta_k+\frac{i}{2}}{u_j-\theta_k -\frac{i}{2}}\right) =\prod_{l\ne j}^M \left( \frac{u_j-u_l+i}{u_j-u_l-i} \right)\period
 \label{bethe}\end{align}
This equation can also be interpreted as a periodicity condition for the phases of the magnon excitations as we go around the chain.
When this equation  is satisfied, the Bethe state is said to be ``on-shell" 
(otherwise called ``off-shell"). 
In that case, the eigenvalue $t_{\mathbf{u}}(u)$ of the transfer 
 matrix $T(u)$ is given by 
\begin{align} t_{\boldsymbol{u} }(u)= \Qt^+(u) \frac{\Qu^{--}(u)}{\Qu(u)}
+\Qt^-(u) \frac{\Qu^{++}(u)}{Q_{\bsu}(u)} 
\comma 
\label{Baxter}
\end{align}
which is sometimes  called the Baxter equation (\ref{bethe}) for the $Q$-function
defined as 
\begin{align}
 \Qu (u)=\prod_{k=1}^M(u-u_k) \period
\end{align}
 The  equation  (\ref{Baxter}) is equivalent 
 to the Bethe ansatz equation through the condition that $t_{\boldsymbol{u}}(u)$ 
 has no poles despite the presence of $Q_{\boldsymbol{u}}(u)$ in the 
 denominator. 

The properties  of the operators $A(u),\ldots ,D(u)$ under the 
 global SU(2) generators $S^i$ are often quite informative. 
For instance, from the transformation properties 
\begin{align}
{}[ S^{z}, B(u)]&=-B(u) \comma  \\
{}[ S^+, B(u)]&= A(u)-D(u) \comma 
\end{align}
one can easily show 
 that if the  Bethe state $|\bsu \rangle$ is on-shell it is the highest weight state with spin $\frac{L}{2}-M$. On the other hand, if it is off-shell,  although having the same spin $\frac{L}{2}-M$,  it is  a  direct 
 sum of states belonging to various representations and is not a highest weight 
 state. 
\subsection{Determinant formulas \label{sec:det}}
The main purpose of this  work is to develop a method of 
 computing the scalar product of the form 
\begin{align}
\langle \bsv  | \bsu \rangle = \langle \uparrow^L | \prod_{i=1}^MC(v_i) \prod_{j=1}^MB(u_j)|\uparrow^L \rangle  
\label{CBprod}
\end{align}
 for SU(2) spin 
chain using the SoV formalism, leading to a new integral representation 
 of such a product.  Traditionally, however, the computation of such a product 
 has been pursued in the framework of the algebraic Bethe ansatz reviewed 
 in the previous subsection. Although the computation is conceptually 
 quite straightforward as one simply needs  
to move  $C(v_i)$'s all the way through $B(u_j)$'s, using the 
exchange algebra,  and act on the pseudovacuum, in practice this procedure 
 produces a multitude  of  terms which grow exponentially in the number of 
 magnons and becomes intractable.  Fortunately, in the case of 
 the product between an on-shell and an off-shell Bethe states, 
Slavnov discovered a much more concise expression in the form of a 
 determinant, which  was to be called Slavnov's determinant 
formula \cite{Slavnov}. 
More recently, various other types of determinant formulas 
 have been developed, which are intimately related to the Slavnov's determinant. 
Since the configuration for which the Slavnov's formula is valid is precisely the one needed for the computation of the 
three point functions in the super Yang-Mills theory, which motivated our study, 
  it is of interest to sketch in advance that how our new formula will be  related, 
 directly or indirectly, to these different variants of determinant formulas. 

As stated above, let us consider the case where either one of the set of 
 rapidities $\bsu$ or $\bsv$ are on-shell. For definiteness, let us take $\bsv$ to be on-shell.  Then 
the original Slavnov's formula computes the scalar product $\langle \bsv|\bsu\rangle$   as a $M \times M$ 
 determinant of the form 
\begin{align}
&\langle \bsv|\bsu\rangle  = \frac{\prod_{i=1}^M\Qt^+(u_i)\Qt^-(v_i)}{\prod_{i<j} (u_i-u_j)(v_j-v_i)}
 \nonumber  \\ 
&\hspace{11pt}\times\det  \left( \frac{1}{u_m-v_n}\left( \prod_{k\ne n}^M(u_m-v_k-i) -\prod_{k\ne n}^M(u_m-v_k+i)\prod_{l=1}^L\frac{u_m-\theta_l -\frac{i}{2}}{u_m-\theta_l +\frac{i}{2}} \right) \right)_{1\leq m,n\leq M} \period \label{eq:first}
\end{align}
Very recently, Kostov and Matsuo \cite{KM}  showed that this expression is equivalent to an alternative determinantal expression of the form 
\begin{align} \langle \bsv|\bsu\rangle = (-1)^M  Z^{\rm{KM}}(\boldsymbol{z} | \boldsymbol{ \theta} )\comma \qquad \boldsymbol{z}\equiv \boldsymbol{u} \cup\boldsymbol{v}\label{eq:KM}
\end{align}
where $Z^{\rm{KM}}(\boldsymbol{z}   |\boldsymbol{ \theta } )$ is 
now a $2M \times 2M$ determinant  given by 
\begin{align}
 Z^{\rm{KM}}(\boldsymbol{z} |\boldsymbol{ \theta } )=
 \frac{\prod_{i=1}^{2M}\Qt^- (z_i)}{\prod_{i<j}(z_i-z_j)}
\det \left( 
 z_m^{n-1}-\prod_{l=1}^{L}\frac{z_m-\theta_l +i/2}{z_m-\theta_l -i/2}(z_m+i)^{n-1}
 \right)_{1\leq m ,n \leq 2M}\period
\label{eq:second} 
\end{align}
They also pointed out that this equivalence is due essentially 
 to the following equality valid when  $\bsu$ or $\bsv$
 are on-shell:
\begin{align}
\langle \uparrow^L | \prod_{i=1}^M C(v_i) \prod_{j=1}^M B(u_j) |\uparrow^L \rangle \propto \langle \downarrow^L | (S^{-})^{L-2M}  \prod_{i=1}^MB(v_i)\prod_{j=1}^M B(u_j) |\uparrow^L \rangle  \period
\label{eq:KM-trick}
\end{align}
Intuitively this can be understood in the following way.
     Suppose 
 the set of rapidities $\boldsymbol{v}$ are on-shell. Then the Bethe state
$\prod_{i=1}^MB(v_i)|\uparrow^L \rangle$ built on the up vacuum 
 is the highest weight state of global SU(2) with spin ${L \over 2} -M$. 
On the other hand, the state $\prod_{i=1}^MC(v_i)|\downarrow^L \rangle$ 
generated by the action of $C(v)$ on 
 the down pseudovacuum has the same eigenvalue for the transfer matrix 
 $T(u)$. Generally,  an on-shell state corresponding to the same solution of the Bethe ansatz equations is  expected to belong to the same SU(2) multiplet.
Since $\prod_{i=1}^MC(v_i)|\downarrow^L \rangle$ is a lowest weight state with spin $-\frac{L}{2}+M$, we can  make it into the highest 
weight state with spin ${L \over 2} -M$ by the action of $(S^+)^{L-2M}$. 
Therefore we should have the equality 
\begin{align}
\prod_{i=1}^MB(v_i)|\uparrow^L \rangle  \propto (S^+)^{L-2M} \prod_{i=1}^MC(v_i)|\downarrow^L \rangle \period
 \label{eq:con}
\end{align}
Taking the conjugate of this relation, we obtain (\ref{eq:KM-trick}).  This identification will be of 
crucial importance when we develop the SoV method for the computation 
 of the scalar product in section \ref{sec-4}. 
 
Now it turns out that our integral formula will be more directly related to 
 another variant of the determinant formula, found by Foda and Wheeler \cite{FW2}. They showed that the Kostov-Matsuo expression 
$Z^{\rm{KM}}(\boldsymbol{z})$ can be identified with 
the so-called partial domain wall partition function  (pDWPF) $Z^{\rm{pDWPF}}(\boldsymbol{z}  |\boldsymbol{ \theta } )$, which naturally arises 
 in the context of the six vertex model:
\begin{align}
Z^{\rm{pDWPF}}(\boldsymbol{z}  |\boldsymbol{ \theta } )&=  \frac{\prod_{\alpha =1}^{2M}\Qt ^{+}(z_\alpha )\Qt ^{-}(z_\alpha ) }{ \prod_{\alpha <\beta } (z_\alpha -z_\beta )\prod_{i<j}(\theta_j -\theta_i)} \nonumber \\
&\times \det \left( \begin{array}{ccc} 
\frac{i}{(z_1 -\theta_1 +i/2)(z_1 - \theta_1 -i/2)}&\cdots &\frac{i}{(z_1 -\theta_L +i/2)(z_1 - \theta_L -i/2)} \\
\vdots & \ \ & \vdots \\
\frac{i}{(z_{2M} -\theta_1 +i/2)(z_{2M} - \theta_1 -i/2)}&\cdots &\frac{i}{(z_{2M} -\theta_L +i/2)(z_{2M} - \theta_L -i/2)} \\
\ \ & \ \ & \ \ \\
\theta_1^{L-2M-1} &\cdots & \theta_L^{L-2M-1} \\
\vdots & \ \ & \vdots \\
\theta_1^0 & \cdots & \theta_L^0 \label{eq:third} 
 \end{array}
 \right) \period 
\end{align}
 It is this expression which will be 
 shown, in the Appendix \ref{ap-b}, to be equivalent to our multiple integral 
 formula. In proving this equality, another more general determinant 
 formula will be of use. It is the Izergin's $L\times L$  determinant \cite{Izergin} expressing
 the domain wall partition function. A slight generalization of our integral 
 formula will be shown to be equal to this Izergin's large determinant 
and by taking the limit where $L-2M$ rapidities are sent to infinity and get 
decoupled, 
 our formula and the Foda-Wheeler formula emerges on the respective side. 
As the latter can be directly shown to be equivalent to the Slavnov determinant
\cite{FW2}, this  proves that our formula is an alternative
 representation of the Slavnov's formula. 
\section{Separation of variables for integrable models \label{sec-3}}
As reviewed in the previous section, excited states in the XXX spin chain $\prod_i B(u_i)\ket{\uparrow^{L}}$ are characterized as a collection of  magnon excitations on top of the ground state and they are distinguished by a set of complex parameters called the Bethe roots, $\{u_i\}$, which are normally interpreted as {\it the rapidities of the magnons}. Then the periodicity condition for such excitations leads to the Bethe equation \eqref{bethe}. In this paper, however, we advocate an alternative view of the excited states, namely that the  states are characterized by the nodes (zeros) of their wave functions and the Bethe roots are interpreted 
instead as {\it the positions of  the nodes}. In this perspective, the Bethe equation arises as a consistency condition for the nodes of the wave function.

To illustrate the basic idea, let us first discuss a simpler example, a one-dimensional harmonic oscillator\fn{This toy model is discussed in a similar manner also in \cite{GV0}}.
As is well-known, the Schr\"{o}dinger equation for the harmonic oscillator can be explicitly solved in terms of the Hermite polynomials. However, here we shall take a slightly different route and try to determine the spectrum without explicitly solving the equation. For this purpose, let us first re-express the Schr\"{o}dinger equation by dividing both sides by the wave function $\psi(x)$: 
\beq{
-\frac{\hbar^2}{2m \psi(x)}\frac{d^2 }{dx^2} \psi (x)+ \frac{m\omega^2 x^2}{2}=E \period\label{HOeq}
}
Then, by studying the behavior of \eqref{HOeq} at large $x$, we conclude that $\psi(x)$ should behave as $\psi(x)\sim \exp \left( -m\omega x^2/2\hbar\right)$ when $x$ is large.  For excited states, $\psi(x)$ must also contain a polynomial prefactor, which gives rise to nodes of the wave functions. Therefore, to characterize $\psi(x)$ by the position of the nodes, let us write down the following ansatz for $\psi(x)$,
\beq{
\psi(x) = \prod_{i=1}^{N} (x-x_i) e^{-m\omega x^2/2\hbar}\period\label{HOwave}
}
Substituting this  ansatz  into \eqref{HOeq}, we obtain the following equation
\beq{
\sum_{i< j}\frac{2}{(x-x_i)(x-x_j)} + \hbar \omega\left( \sum_i\frac{ x}{x-x_i} + \frac{1}{2}\right)=E\period \label{HObax}
}
Then from its large $x$ behavior, the energy $E$ is determined in terms of  the number of nodes as $E=\hbar \omega (N+1/2)$. In addition, since the RHS of \eqref{HObax} is a constant and free of poles, we must demand that the residue of the poles at $x=x_i$ on the LHS must vanish. This leads to  a Bethe-ansatz-like equation for the positions of the nodes of the wave function,
\beq{
x_i = \frac{\hbar}{2m\omega}\sum_{j\neq i} \frac{1}{x_i-x_j}\period\label{HObethe}
}

Although this  idea of characterizing the excited states in terms of the number and the positions of the nodes is quite elementary and intuitive, it is technically difficult to apply this idea directly to the system with many degrees of freedom. However, in the case of the integrable models, it is often possible to decompose the system into a set of mutually decoupled one dimensional problems. The systematic method to carry this out is the method of separation of variables developed by Sklyanin, which we will explain in the rest of this section. By applying this method, we will see explicitly in section \ref{sec-4} that the Bethe equation for the XXX spin chain can indeed be interpreted as a consistency equation for the nodes of the wave function as in \eqref{HObethe}. 
\subsection{Basic notions of the separation of variables}
Before delving into the details of the method developed by Sklyanin, here we briefly summarize the basic notion of the separation of variables. 
In classical mechanics, separation of variable is applicable only  if there are as many number of conserved charges, $h_1,\ldots,h_d $, as the dynamical variables. In such a case, by a judicious choice of canonical variables, it is often possible to write down a set of equations, each of which contains only one canonical pair $\{x_k ,p_k\}$:
 \beq{
 W_k (x_k ,p_k ;h_1,\ldots ,h_d)=0\comma \qquad k=1\ldots d\period\label{sov-cl}
}
This type of equation is analogous to the expression of the energy of the harmonic oscillator, $E=p^2/2m + m\omega^2 x^2 /2$, and one can determine the classical motion of the system in much the same way as in that case. 

When we consider the quantum system, the equations \eqref{sov-cl} are replaced by the following equations for the eigenstates of the conserved charges,
\beq{
W_k (\hat{x}_k ,\hat{p}_k ;h_1,\ldots ,h_d)\ket{\Psi}=0\comma \qquad k=1\ldots d\period\label{sov-op}
}
In terms of the wave function in the coordinate representation, $\Psi(x_1,\ldots ,x_d)$, \eqref{sov-op} can be re-expressed as
\beq{
W_k \left( x_k , \frac{\hbar}{i}\frac{\del}{\del x_k} ;h_1,\ldots ,h_d\right)\Psi(x_1,\ldots ,x_d)=0\comma \qquad k=1\ldots d\period\label{sov-wv}
}
It is easy to see that \eqref{sov-wv} admits a completely factorized solution, $\Psi=\prod_k\psi_k(x_k)$, each factor of which satisfies the following one dimensional equation, 
\beq{
W_k \left( x_k , \frac{\hbar}{i}\frac{\del}{\del x_k} ;h_1,\ldots ,h_d\right) \psi_k(x_k) =0\period
}
In this way, the original system with many degrees of freedom can be reduced to a set of mutually decoupled one dimensional systems.
\subsection{Sklyanin's magic recipe}
The most nontrivial step in the procedure  above is the construction of the 
separated variables satisfying the  equations of the form 
\eqref{sov-cl} or \eqref{sov-op}. This is indeed a difficult problem for interacting many-body systems. However, for the integrable models which can be formulated in terms of the Lax operators, Sklyanin proposed a systematic method for the construction, often  referred to as the Sklyanin's magic recipe. In what follows, we sketch the essence of this  recipe\fn{The discussion here is basically restricted to the simplest class of the integrable models, called {\it rational} models. For {\it trigonometric} or {\it elliptic} models, nontrivial modification of the method is required \cite{Sklyanin}.} applied to systems with a $2\times 2$ monodromy matrix. More precise  analysis for the case of the XXX spin chain will be given in  the next section.

For simplicity, let us first consider the classical case. In a classically integrable system with a $2\times 2$ monodromy matrix 
\beq{
\Omega (u)=\pmatrix{cc}{A(u)&B(u)\\C(u)&D(u)}\comma
}
there is an immediate candidate for the set of equations \eqref{sov-cl}. 
 It is the {\it characteristic equation} for the monodromy matrix
\beq{
\det \left( z-\Omega (x)\right) =0\comma\label{ch}
}
where $z$ is the eigenvalue of the matrix $\Omega(x)$.
Since the expansion of $\Omega(x)$ in powers of $x$ yields a set of conserved charges as its coefficients, \eqref{ch} is indeed of the form of \eqref{sov-cl} if we can somehow identify $x$ and $z$ with dynamical variables. The recipe proposed by Sklyanin is to use the solutions $x_k$'s to the equation $B(u)=0$ as $x$-variables:
\beq{
B(u)=(u-x_1)(u-x_2)\cdots\period
}
In the case of the lattice models, such as the XXX spin chain discussed in the previous section, $B(u)$ is a polynomial in $u$, whose order basically  equals  the lattice size. Therefore, this prescription indeed provides the  correct number of variables. Furthermore, owing to the Poisson commutativity among  $B(u)$'s,   $x_k$'s also 
commute with each other and thus they are mutually independent separated variables. On the other hand, the $z$-variables, which are the eigenvalues of $\Omega(x)$, are provided by the diagonal components, $A(x_k)$ or $D(x_k)$, since $\Omega(x_k)$ becomes a lower triangular matrix owing to $B(x_k)=0$. Then the remaining task is to understand the relation of $A(x_k)$ and $D(x_k)$ to  the conjugate momenta  $p_k$, which satisfy the standard  commutation relations:
\beq{
\{x_k ,x_l\}=0\comma \qquad \{p_k ,p_l\}=0 \comma \qquad \{x_k,p_l\}=\delta_{kl}\period
}
In most cases, by explicitly computing the Poisson brackets of $A(x_k)$ and $D(x_k)$ with $x_k$, we can show  that they are related to $p_k$ roughly as 
\beq{
 A(x_k) \sim e^{ip_k}\comma \qquad D(x_k) \sim e^{-ip_k}\period\label{ADp} 
}

In the case of the quantum integrable models, separated variables $x_k$'s become a set of commuting operators $\hat{x}_k$'s, which  are characterized as the roots of the {\it operator} equation $B(u)=0$. Just as for the classical case, the conjugate operators, $e^{i\hat{p}_k}$ and $e^{-i\hat{p}_k}$, are given\fn{Note, in the quantum case, we need to consider the ordering of the operators. In the case of the XXX spin chain, this is explicitly worked out in section \ref{sec-3}.} essentially by $A(\hat{x}_k)$ or $D(\hat{x}_k)$. To derive a set of one dimensional equations of the type  \eqref{sov-wv}, let us consider the wave function in the $x_k$-basis:
\beq{
\Psi( x_1,\ldots ,x_d)=\braket{x_1,\ldots,x_d}{\Psi}\comma
}
where $\bra{x_1,\ldots,x_d}$ is an eigenstate of the operators $\hat{x}_k$'s. Now if the state $\ket{\Psi}$ is an eigenstate of $T(u) = A(u) +D(u)$, a generating function of the commuting set of Hamiltonians, we can  compute $\bra{x_1,\ldots,x_d}T(\hat{x}_k)\ket{\Psi}$ as
\beq{
\bra{x_1,\ldots, x_d}T(\hat{x}_k)\ket{\Psi} = t(x_k)\Psi (x_1,\ldots ,x_d)\comma\label{tP}
}
where $t(u)$ is the eigenvalue of $T(u)$ for $\ket{\Psi}$.
We can evaluate the same quantity  also by acting $T(\hat{x}_k)$ to the left on $\bra{x_1,\ldots,x_d}$. To carry this out we use the relation of  $T(\hat{x}_k)$ with the momenta $\hat{p}_k$, \ie 
\beq{
T(\hat{x}_k)=A(\hat{x}_k)+D(\hat{x}_k)\sim e^{i\hat{p}_k}+e^{-i\hat{p}_k}
\period 
}
Then we find 
\beq{
\bra{\ldots, x_k,\ldots}T(\hat{x}_k) \sim\bra{\ldots, x_k+1,\ldots}+ \bra{\ldots, x_k-1,\ldots}\period
}
In this way we arrive at the following equation for the wave function $\Psi$:
\beq{
t(x_k)\Psi (\ldots , x_k ,\ldots )\sim\Psi(\ldots , x_k +1,\ldots ) +\Psi(\ldots , x_k -1,\ldots )\comma \quad k=1 ,\ldots ,d\period\label{tP-gen}
}
Assuming the  factorized form of the wave functions  $\Psi (x_1,\ldots,x_d)=\prod_k \psi_k(x_k)$, we can decompose \eqref{tP-gen} into a set of mutually decoupled one dimensional equations:
\beq{
t(x_k) \psi_k (x_k)\sim\psi_k (x_k +1) +\psi_k (x_k-1)\period\label{tq-gen}
}
This equation is the analogue of the Schr\"{o}dinger equation for 
 the harmonic oscillator. Therefore, as in that case, we can derive a consistency condition for the nodes of the wave function. Assuming a form of $\psi_k$ as $\psi_k(x)=\prod_l (x-u_l)$ and setting $x_k=u_j$ in \eqref{tq-gen}, we obtain the algebraic relations for the positions $u_j$ of the nodes:
 \beq{
 1\sim \prod_{l\neq j} \frac{u_j-u_l+1}{u_j-u_l-1}\period\label{beq-gen}
 }
 Note that this  is identical with  the Bethe equation. Therefore, as mentioned at the beginning of this section, the Bethe roots can be interpreted as the nodes of the wave function in this approach. In the next section, we will see that the logic outlined here is explicitly realized in the case of the XXX spin chain.
 
\section{Integral representation of the scalar products for XXX spin chain\label{sec-4}}
In the preceding section, we sketched the basic idea of the method of separated variables for integrable models. In this section, we will apply it to the periodic SU(2) XXX spin chain and derive a multiple integral representation of the scalar products
 in the basis where the separated variables are diagonal. 
The resultant expression can be brought to a form which resembles the integral over the eigenvalues of a matrix model. 
\subsection{Construction of the separated variables}
Recall the definition of the monodromy matrix  $\Omega(u)$ 
for the XXX spin chain with inhomogeneity parameters $\theta_k$:
\beq{
&\Omega (u) = \pmatrix{cc}{A(u)&B(u)\\C(u)&D(u)}\equiv L_1(u-\theta_1 )L_2(u-\theta_2 )\cdots L_L(u-\theta_L )\comma\\
&L_{k}(u)\equiv \pmatrix{cc}{u+iS_{k}^{z}&iS_{k}^{-}\\iS_{k}^{+}&u-iS_{k}^{z}}\period
}
As outlined in the previous section, the separated variables for  integrable models with a $2\times 2$ monodromy matrix are usually given by the roots of the operator equation, $B(u)=0$. However, as already pointed out in the introduction, 
 in the case of the periodic SU(2) spin chain,  the operator $B(u)$ is proportional to $S^{-}$ in the large $u$ limit as  $B(u)\sim iS^{-}u^{L-1}+\cdots$, and is not diagonalizable. This problem can be circumvented by introducing a  twisting matrix $K_{\epsilon}=\pmatrix{cc}{1&\epsilon\\-\epsilon &1}$, which 
 changes the boundary condition  and modifies  the monodromy matrix as
\beq{
\Omega_{\epsilon}(u)=K_{\epsilon}\Omega(u)\equiv \pmatrix{cc}{A_{\epsilon}(u)&B_{\epsilon}(u)\\C_{\epsilon}(u)&D_{\epsilon}(u)}\period\label{twisting}
}
Although such a  twisting changes the dynamical properties of the spin chain, it does not affect the computation of the scalar products since, 
 as we shall show explicitly later in this section,  they  can 
 be re-expressed in terms of quantities independent of the twisting parameter $\epsilon$.  After twisting, the large $u$ behavior of $B_{\epsilon}(u)$ is modified to $B_{\epsilon}(u) \sim \epsilon u^{L}+i(S^{-}-\epsilon S^{z}
+ i \sum_{j}\theta_j )u^{L-1}+\cdots$ and $B_{\epsilon}(u)$ becomes diagonalizable. Then it can be factorized as
\beq{
B_{\epsilon}(u)=\epsilon \prod_{k=1}^{L}(u-\hat{x}_k)\comma
} 
where $\hat{x}_k$'s are the roots of the operator equation, $B_{\epsilon}(u)=0$. As the twisting preserves the algebra among the elements $A(u),\cdots ,D(u)$, the operators $B_{\epsilon}$'s  continue to commute with each other, namely $[B_{\epsilon }(u),B_{\epsilon }(v)]=0$, and this implies that  $\hat{x}_k$'s 
also mutually commute: 
$[\hat{x}_k\comma \hat{x}_l]=0$.
These operators are the ``coordinates" of the separated variables and one can consider their left eigenstates and right eigenstates, $\bra{x_1,\ldots,x_L}$ and $\ket{x_1,\ldots,x_L}$, upon which $B_{\epsilon}(u)$ acts in the following way:
\beq{
\bra{x_1,\ldots ,x_L} B_{\epsilon}(u)=\left( \epsilon \prod_{k=1}^{L}(u-x_k)\right)\bra{x_1,\ldots ,x_L}\comma\label{bl}\\
B_{\epsilon}(u)\ket{x_1,\ldots,x_L}=\left( \epsilon \prod_{k=1}^{L}(u-x_k)\right)\ket{x_1 ,\ldots ,x_L}\period\label{br}
}
As  explained in the Appendix \ref{ap-a}, the eigenvalue of the operator $\hat{x}_k$ 
takes only two values given by 
$\theta_k \pm \frac{i}{2}$. As a consequence, the dimension of the Hilbert space spanned by the eigenstates of the separated variables is $2^{L}$, which precisely matches that of the spin chain Hilbert space. This assures the completeness of the separated variable basis.

At $\xhat_k$ the operator $B_\ep$ vanishes and  the form of the monodromy matrix   becomes lower triangular. Therefore  the two eigenvalues are given 
by  $A_{\epsilon}(\hat{x}_k)$ and  $D_{\epsilon}(\hat{x}_k)$, which 
 are expected to be identified as $e^{\pm i\hat{p}_k}$, where $\hat{p}_k$ 
 is  the momentum operator conjugate to $\hat{x}_k$. 
To see this more precisely, since $A_{\epsilon}(u)$ and $D_{\epsilon}(u)$ are polynomials in $u$ with operator-valued coefficients, we need to specify the ordering of  $\hat{x}_k$ and  the coefficients, which are also operators in general. 
The ordering appropriate for the left eigenstates, to be denoted by  ${\rm :}\ast\ast\ast {\rm :}_{L}$, turns out to be placing  all the $\hat{x}_k$'s to the left of the coefficients, namely
\beq{ {\rm :}F(\hat{x}_k){\rm :}_L&\equiv \sum _n \hat{x}_k^n \hat{F}_{n}
\comma \qquad \quad \text{for}\quad F(u)= \sum _n u^n \hat{F}_{n}\period
}
Then the commutation relation between $A_{\epsilon}(u)$ and 
$B_{\epsilon}(u)$, given by  $(u-v)A_{\epsilon}(v)B_{\epsilon}(u)=(u-v +i)B_{\epsilon}(u)A_{\epsilon}(v)-iB_{\epsilon}(v)A_{\epsilon}(u)$, leads to
\beq{
{\rm :}(u-\hat{x}_k)A_{\epsilon}(\hat{x}_k)B_{\epsilon}(u){\rm :}_{L}=\,{\rm :}(u-\hat{x}_k +i)B_{\epsilon}(u)A_{\epsilon}(\hat{x}_k){\rm :}_{L}-i{\rm :}B_{\epsilon}(\hat{x}_k)A_{\epsilon}(u){\rm :}_{L}\period\label{AB-com}
}
Since the second term on  the RHS of \eqref{AB-com}, 
 containing $B_{\epsilon}(\hat{x}_k)$,  vanishes,  and since  $B_{\epsilon}(u)$ commutes with $\hat{x}_k$, we can simplify \eqref{AB-com} to
\beq{
(u-\hat{x}_k){\rm :}A_{\epsilon}(\hat{x}_k){\rm :}_L B_{\epsilon}(u)=(u-\hat{x}_k+i)B_{\epsilon}(u){\rm :}A_{\epsilon}(\hat{x}_k){\rm :}_L\comma \label{AB-com2}
}
where the normal-ordering is now imposed  only on $A_{\epsilon}(\hat{x}_k)$. 
Then by acting \eqref{AB-com2} to the left eigenstate, we obtain
\beq{
(u-x_k)\bra{x_1 ,\ldots ,x_L}{\rm :}A_{\epsilon}(\hat{x}_k){\rm :}_L  B_{\epsilon}(u)=\epsilon (u-x_k+i)\prod_{l=1}^{L}(u-x_l) 
\bra{x_1 ,\ldots ,x_L}{\rm :}A_{\epsilon}(\hat{x}_k){\rm :}_L
\period\nn
}
Dividing both sides  by $(u-x_k)$, we see that $B(u)$ acting on 
 the state $\bra{x_1 ,\ldots ,x_L}{\rm :}A_{\epsilon}(\hat{x}_k){\rm :}_L$ vanishes at $u=x_k-i$. This means that the operator ${\rm :}A_{\epsilon}(\hat{x}_k){\rm :}_L$ indeed effects the shift of  the eigenvalue of $\hat{x}_k$ by $-i$, 
 namely\footnote{For the literal identification of ${\rm :}A_{\epsilon}(\hat{x}_k){\rm :}_L$ with $e^{i\hat{p}_k}$, it is more natural to  rename $\xhat_k$
 as $-i \xhat_k$. Then, the new $\xhat_k$ gets shifted by $+1$ and 
its spectrum becomes real at $\theta_k=0$. But we shall not do this and stick to 
 the customary definition.}
\beq{
\bra{\ldots, x_k ,\ldots}{\rm :}A_{\epsilon}(\hat{x}_k){\rm :}_L\propto \bra{\ldots, x_k -i,\ldots}\period\label{A-prop}
}
A similar argument for $D_{\epsilon}(u)$ leads to the conclusion that ${\rm :}D_{\epsilon}(\hat{x}_i){\rm :}_L$ shifts the eigenvalue of $\hat{x}_k$ by $+i$,
\beq{
\bra{\ldots, x_k ,\ldots}{\rm :}D_{\epsilon}(\hat{x}_k){\rm :}_L\propto \bra{\ldots, x_k +i,\ldots}\period\label{D-prop}
}
The  constants of proportionality in \eqref{A-prop} and \eqref{D-prop} 
can  be determined by the analysis detailed in the Appendix \ref{ap-a}. 
 Since these results, together with the spectrum of $\xhat_k$ already quoted, 
 are  basic to the rest of the analysis, we shall display them as a
 theorem:
\\
\underline{\it Theorem 1}:\\

(i) \quad The spectrum of $\hat{x}_k$ is given by the two values\footnote{
As shown in Appendix \ref{ap-a}, what one can show is that 
 the spectrum of each $\hat{x}_j$ is of the form 
$\theta_k\pm {i \over 2}$ for some $k$. 
 Here and hereafter we adopt the natural convention to associate 
 the spectrum $\theta_k\pm {i \over 2}$ with $\hat{x}_k$. 
 }
\begin{align}
x_k &= \theta_k +  {i \over 2} \comma \hspace{11pt} \theta_k-{i \over 2}. 
\end{align}

(ii)\quad 
The operators 
${\rm :}A_{\epsilon}(\hat{x}_k){\rm :}_L$ and ${\rm :}D_{\epsilon}(\hat{x}_k){\rm :}_L$ act on the left eigenstates in the following manner
\beq{
\bra{\ldots, x_k ,\ldots}{\rm :}A_{\epsilon}(\hat{x}_k){\rm :}_L=\sqrt{1+\epsilon^2}\Qt ^{+}(x_k) \bra{\ldots, x_k -i,\ldots}\comma\label{bra-A}\\
\bra{\ldots, x_k ,\ldots}{\rm :}D_{\epsilon}(\hat{x}_k){\rm :}_L=\sqrt{1+\epsilon^2}\Qt ^{-}(x_k) \bra{\ldots, x_k +i,\ldots}\period\label{bra-D}
}

For the right eigenstates, an appropriate ordering prescription is to put all $x_k$'s to the right of the coefficients of $A_{\epsilon}(u)$ and $D_{\epsilon}(u)$:
\beq{
&{\rm :} F(\hat{x}_k){\rm :}_R\equiv \sum _n \hat{F}_{n}\hat{x}_k^n \comma\qquad \qquad\text{for}\quad F(u)= \sum _n u^n \hat{F}_{n} \period
} 
Then the action of ${\rm :}A_{\epsilon}(\hat{x}_k){\rm :}_R$ and ${\rm :}D_{\epsilon}(\hat{x}_k){\rm :}_R$ on the right eigenstates are expressible as
\beq{
{\rm :}A_{\epsilon}(\hat{x}_k){\rm :}_R\ket{\ldots, x_k ,\ldots}=\sqrt{1+\epsilon^2}\Qt ^{-}(x_k) \ket{\ldots, x_k +i,\ldots}\comma\label{ket-A}\\
{\rm :}D_{\epsilon}(\hat{x}_k){\rm :}_R\ket{\ldots, x_k ,\ldots}=\sqrt{1+\epsilon^2}\Qt ^{+}(x_k) \ket{\ldots, x_k -i,\ldots}\period\label{ket-D}
}
Since $A_{\epsilon}(u)$ and $D_{\epsilon}(u)$ are $L$-th order polynomials in $u$ with a unit leading coefficient, the action of these operators at $L$ distinct values of $u$, \eqref{bra-A} and \eqref{bra-D}, completely determines the explicit forms of the operators as follows:
\beq{
A_{\epsilon}(u) = \prod_{k=1}^{L} (u-\hat{x}_k) + \sum_{k=1}^{L}\left( \prod_{j\neq k} \frac{u-\hat{x}_j}{\hat{x}_k-\hat{x}_j}\right) {\rm :}A_{\epsilon}(\hat{x}_k){\rm :}_{L}\comma\label{Au-L}\\
D_{\epsilon}(u) = \prod_{k=1}^{L} (u-\hat{x}_k) + \sum_{k=1}^{L}\left( \prod_{j\neq k} \frac{u-\hat{x}_j}{\hat{x}_k-\hat{x}_j}\right) {\rm :}D_{\epsilon}(\hat{x}_k){\rm :}_L\period\label{Du-L}
} 
They are expressible also in terms of the right-ordered operators, ${\rm :}A_{\epsilon}(\hat{x}_k){\rm :}_R$ and ${\rm :}D_{\epsilon}(\hat{x}_k){\rm :}_R$, as
\beq{
A_{\epsilon}(u)&=\prod_{k=1}^{L} (u-\hat{x}_k) + \sum_{k=1}^{L} {\rm :}A_{\epsilon}(\hat{x}_k){\rm :}_R\left( \prod_{j\neq k} \frac{u-\hat{x}_j}{\hat{x}_k-\hat{x}_j}\right)\comma\label{Au-R}\\
D_{\epsilon}(u)&=\prod_{k=1}^{L} (u-\hat{x}_k) + \sum_{k=1}^{L} {\rm :}D_{\epsilon}(\hat{x}_k){\rm :}_R\left( \prod_{j\neq k} \frac{u-\hat{x}_j}{\hat{x}_k-\hat{x}_j}\right)\period\label{Du-R}
}

From \eqref{Au-L} and \eqref{Du-L}, we can derive a difference equation for the eigenstate $\ket{\psi}$ of the (twisted) transfer matrix, $T_{\epsilon}(u)\equiv A_{\epsilon}(u)+D_{\epsilon}(u)$. This is done by computing $\bra{x_1,x_2 ,\ldots, x_L}T_{\epsilon}(u)\ket{\psi}$ in two different ways: First by acting $T_{\epsilon}(u)$ on $\bra{x_1,x_2 ,\ldots, x_L}$ using \eqref{Au-L} and \eqref{Du-L}, and second by acting it on $\ket{\psi}$. By setting $u=x_k$ in the resulting equation, we obtain the following simple equation for the wave function of the eigenstate, $\Psi(x_1 ,\ldots ,x_L)\equiv \langle x_1 ,\ldots,x_L|\psi\rangle$:
\beq{
\frac{t_\ep(x_k)}{\sqrt{1+\epsilon^2}}\Psi(\ldots ,x_k,\ldots) =\Qt ^{+} (x_k) \Psi (\ldots ,x_k-i,\ldots) +\Qt ^{-} (x_k) \Psi (\ldots ,x_k+i,\ldots)\period\label{tq1}
}
Here  $t_{\epsilon}(u)$ is the eigenvalue of $T_{\epsilon}(u)$, \ie $T_{\epsilon}(u)\ket{\psi}=t_{\epsilon}(u)\ket{\psi}$.
Assuming a factorized form of the wave function, $\Psi (x_1 ,x_2,\ldots ,x_L )=\psi_1(x_1)\psi_2 (x_2)\ldots \psi_L(x_L)$, \eqref{tq1} can be decomposed into a set of $L$ one-dimensional equations, which can be regarded as the ``Schr\"{o}dinger equations" for the separated variables:
\beq{
\frac{t_{\epsilon}(x_k)}{\sqrt{1+\epsilon^2}} \psi_k(x_k)=\Qt ^{+} (x_k) \psi_k^{--}(x_k)+\Qt ^{-} (x_k) \psi_k^{++}(x_k)\period\label{sov-bax}
}
In the $\epsilon \to 0$ limit, the equation \eqref{sov-bax} for $\psi_k$ apparently 
takes the same form as the Baxter equation \eqref{Baxter} for the $Q$-function, $\Qu (u)$. However one should keep in mind that $\Qu$ and $\psi_k(x_k)$ are conceptually quite different: While $\Qu$ is introduced as a polynomial with zeros at the rapidities of the magnon excitations and can be defined on the whole complex plane, $\psi_k$ is the wave function in the separated variable basis and is defined only on the discrete eigenvalues  $x_k =\theta_k \pm i/2$. Therefore it is {\it a priori} not clear whether we can identify $\psi_k$ with $\Qu$. Nevertheless, as we shall  later see explicitly, the factor representing the wave function in the multiple integral formula is given indeed by the $Q$-function. Therefore, as far as the multiple integral formula is concerned, $\psi_k$ can be  identified with $\Qu$ and the Bethe equation  can be interpreted as the consistency condition for the zeros of the wave function. This  evidently  parallels  the case of the harmonic oscillator d
 iscussed in section \ref{sec-3}. 
 
\subsection{Multiple integral representation for scalar products}
Having constructed the separated variables, our next goal is to express the scalar product between an off-shell and an on-shell Bethe states given by 
 $\langle \text{\boldmath $v$}|\text{\boldmath $u$}\rangle=\bra{\uparrow^{L}}\prod_{i=1}^{M}C(v_i)\prod_{i=1}^{M}B(u_i)\ket{\uparrow^{L}}$ 
as the overlap between two wave functions of separated variables. Our basic strategy for deriving  such a expression is to insert into the scalar product a resolution of unity in the SoV basis, namely
\beq{
\id = \sum_{x_k= \theta_k \pm i/2}\mu(\boldsymbol{x}) \,\ket{\boldsymbol{x}}\bra{\boldsymbol{x}}\comma\label{summation}
}
where $\boldsymbol{x}$ stands for $\{x_1,\ldots,x_L\}$ and $\mu(\boldsymbol{x})$ is the measure factor for the  summation,  to be specified later. Unfortunately, this procedure cannot be carried out straightforwardly 
because  the scalar product of our interest contains the operator 
$C(u)$ and its action on the $B$-diagonal SoV basis is quite complicated. 
In addition, to employ the SoV basis, we need to introduce the twist in the boundary condition as in \eqref{twisting}, which is not present in the original 
scalar product as above. 

The first problem can be circumvented by the trick due to Kostov and Matsuo \cite{KM}, which converts $C(v_i)$  to $B(v_i)$ within the scalar product provided $v_i$'s satisfy the Bethe equation. 
Although not explicitly given in \cite{KM}, one can work out the 
precise factors in the conversion formula and obtain the expression
\beq{
 \bra{\uparrow^{L}}\prod_{i=1}^{M}C(v_i)\prod_{j=1}^{M}B(u_j)\ket{\uparrow^{L}}=&\frac{(-1)^{M}}{(L-2M)!}\bra{\downarrow^{L}}(S^{-})^{L-2M}\prod_{i=1}^{M}B(v_i)\prod_{j=1}^{M}B(u_j)\ket{\uparrow^{L}}\comma\label{KMexact}
} 
which contains only the operator $B(u)$.  This rewriting has another gratifying feature: It allows us to introduce the twist of the boundary condition without changing the value of the scalar product. This is done by replacing the second line in \eqref{KMexact} with
\beq{
\bra{\downarrow^{L}}(S^{-}-\epsilon S^{z}
+ i \sum_{l=1}^{L} \theta_l)^{L-2M}\prod_{i=1}^{M}B_{\epsilon}(v_i)\prod_{j=1}^{M}B_{\epsilon}(u_j)\ket{\uparrow^{L}}\period\label{mproduct}
}
Although \eqref{mproduct} has apparent dependence on $\epsilon$ as well as an extra dependence on $\theta_l$'s, 
such unwanted terms actually vanish\fn{To see this, it suffices  to recall that $B_{\epsilon}(u)$ is composed of $B(u)+\epsilon D(u)$ and $S^-$
 and that $B(u)$ lowers the eigenvalue of $S^z$ by $1/2$  while $S^z$ and $D(u)$ leave it unchanged.} thanks to the conservation of the total spin $S^z$ along the $z$-axis.

Then, inserting a resolution of unity \eqref{summation} into \eqref{mproduct} and using the action of $B_{\epsilon}(u)$ on the SoV basis \eqref{bl} and \eqref{br}, we obtain the following expression\fn{Note that the combination $S^{-}-\epsilon S^{z}+i\sum_j \theta_j$   
appears in  $B_{\epsilon}(u)$ as $B_{\epsilon}(u) \sim \epsilon u^{L}+i(S^{-}-\epsilon S^{z}+i\sum_j \theta_j)u^{L-1}+\ldots$ and its action on the SoV basis is thus given by  $(S^{-}-\epsilon S^{z}+i\sum_j \theta_j)\ket{x_1,\ldots ,x_L} = \epsilon \sum_{i}x_i \ket{x_1,\ldots ,x_L}$.}
\beq{
&\bra{\downarrow^{L}}(S^{-}-\epsilon S^{z}+ i \sum_{l=1}^{L} \theta_l)^{L-2M}\prod_{i=1}^{M}B_{\epsilon}(v_i)\prod_{j=1}^{M}B_{\epsilon}(u_j)\ket{\uparrow^{L}}\nn\\
&=\sum_{x_k= \theta_k \pm i/2}\!\!\epsilon ^L\mu (\boldsymbol{x}) f_{L}(\boldsymbol{x})f_{R}(\boldsymbol{x})\left( \sum_{j=1}^{L}x_j\right)^{L-2M} \!\!\!\prod_{k=1}^{L}\Qu (x_k)\Qv (x_k)\comma\label{sum-rep}
}
where $f_{L,R}$ are given by 
\beq{
f_{L}(\boldsymbol{x})\equiv \langle \downarrow^{L}|\boldsymbol{x}\rangle\comma \qquad f_{R}(\boldsymbol{x})\equiv\langle \boldsymbol{x}| \uparrow^{L}\rangle\period
}
Note that both the measure $\mu(\boldsymbol{x})$ and the functions 
$f_{L,R}(\boldsymbol{x})$  depend on the twist parameter $\epsilon$ but the total expression \eqref{sum-rep} should be $\epsilon$-independent as argued  above.

Let us now determine $\mu(\boldsymbol{x})$ and $f_{L,R}(\boldsymbol{x})$.  To determine $\mu(\boldsymbol{x})$, we consider the overlap\fn{Note $\langle \boldsymbol{x^{\prime}}|\boldsymbol{x}\rangle$ cannot be regarded as a norm since the left and the right eigenstates are not Hermitian conjugate to each other. Therefore $\langle \boldsymbol{x^{\prime}}|\boldsymbol{x}\rangle$ can be in general complex-valued.} between the left and the right eigenstates in the SoV basis, $\langle \boldsymbol{x^{\prime}}|\boldsymbol{x}\rangle$. First, since $\langle \boldsymbol{x^{\prime}}|$ and $|\boldsymbol{x}\rangle$ are both eigenstates of the operators $\hat{x}_k$'s, the overlap vanishes unless the eigenvalues coincide. Therefore, we conclude that $\langle \boldsymbol{x^{\prime}}|\boldsymbol{x}\rangle$ is proportional to $\delta_{x^{\prime}_1,x_{1}}\delta_{x^{\prime}_2,x_{2}}\ldots\delta_{x^{\prime}_L,x_{L}}$. Second, when we act the right hand side of \eqref{summation} on the state $\langle \boldsymbol{x^{\prime}}|$, the state should not change as the left hand side of \eqref{summation} is just an identity operator. Owing to this condition, we can express $\langle \boldsymbol{x^{\prime}}|\boldsymbol{x}\rangle$ in terms of the measure factor $\mu(\boldsymbol{x})$ as
\beq{
\langle \boldsymbol{x^{\prime}}|\boldsymbol{x}\rangle = \mu^{-1}(\boldsymbol{x}) \,\delta_{x^{\prime}_1\comma x_1}\ldots\delta_{x^{\prime}_L\comma x_L}\period\label{norm}
}
This suggests that $\mu(\boldsymbol{x})$ can be determined by 
computing the matrix element 
\beq{
\bra{\boldsymbol{x^{\prime}}}A_{\epsilon}(u)\ket{\boldsymbol{x}}\label{xAx}
}
in two different ways: First, by acting $A_{\epsilon}(u)$ on the bra using \eqref{bra-A} and \eqref{Au-L} and setting $x^{\prime}_j =x_j$ for $j\neq k$ and $x^{\prime}_k=x_k+i$, we obtain 
\beq{
\mu^{-1} (\ldots\comma x_k\comma \ldots)\left( \prod_{j\neq k}\frac{u-x_j}{x_k-x_j+i}\right) \Qt^{+++}(x_k)\period\label{measure1}
}
Second, by acting $A_{\epsilon}(u)$ on the ket using \eqref{ket-A} and \eqref{Au-R} and setting $x^{\prime}_j =x_j$ for $j\neq k$ and $x^{\prime}_k=x_k+i$, we obtain
\beq{
\mu^{-1}(\ldots\comma x_k+i\comma \ldots)\left( \prod_{j\neq k}\frac{u-x_j}{x_k-x_j}\right) \Qt^{-}(x_k)\period\label{measure2}
}
By equating \eqref{measure1} and \eqref{measure2}, we 
arrive at the following recursion relation for $\mu(\boldsymbol{x})$:
\beq{
\frac{\mu(\ldots, x_k+i,\ldots )}{\mu(\ldots ,x_k, \ldots )} =\frac{\Qt^{-}(x_k)}{\Qt^{+++}(x_k)}\prod_{j\neq k}\frac{x_k-x_j+i}{x_k-x_j}\period\label{rec-mu}
}
The solution to this equation can be obtained as
\beq{
\mu(\boldsymbol{x})\propto \prod_{i<j}(x_i-x_j) \prod_{k} e^{-\pi (x_k-\theta_k)}\prod_{l\neq m} \frac{1}{(x_l-\theta_m+\frac{i}{2})(x_l-\theta_m-\frac{i}{2})}\period\label{sum-mu}
}
Similarly, we can derive the recursion relations for $f_{L,R}(\boldsymbol{x})$ by computing $\bra{\downarrow^{L}}A_{\epsilon}(u)\ket{\boldsymbol{x}}$ and $\bra{\boldsymbol{x}}A_{\epsilon}(u)\ket{\uparrow^{L}}$ in two different ways. First, by acting  $A_{\epsilon}(u)$ on the bra using 
the formula,
\beq{
\bra{\downarrow^{L}}A_{\epsilon}(u) =\bra{\downarrow^{L}} \left( A(u)+\epsilon C(u)\right) =\Qt^{-}(u)\bra{\downarrow^{L}}\comma
}
and \eqref{Au-L}, we obtain
\beq{
\bra{\downarrow^{L}}A_{\epsilon}(u)\ket{\boldsymbol{x}}=&\Qt^{-}(u)f_{L}(\boldsymbol{x})\comma\label{flr1}\\
\bra{\boldsymbol{x}}A_{\epsilon}(u)\ket{\uparrow^{L}}=&\prod_{k=1}^{L}(u-x_k)f_{R}(\boldsymbol{x})\nn\\
&+\sqrt{1+\epsilon^2}\sum_{k=1}^{L}\left( \prod_{j\neq k}\frac{u-x_j}{x_k-x_j}\right) \Qt^{+}(x_k)f_{R}\left( \ldots\comma x_k-i\comma \ldots\right)\period\label{flr2}
}
Second, by acting $A_{\epsilon}(u)$ on the ket using \eqref{Au-R} and the formula
\beq{
 A_{\epsilon}(u)\ket{\uparrow^{L}}=\left( A(u)+\epsilon C(u)\right)\ket{\uparrow^{L}} =\Qt^{+}(u)\ket{\uparrow^{L}}\comma
}
we obtain
\beq{
\bra{\downarrow^{L}}A_{\epsilon}(u)\ket{\boldsymbol{x}}=&\prod_{k=1}^{L}(u-x_k)f_{L}(\boldsymbol{x})\nn\\&+\sqrt{1+\epsilon^2}\sum_{k=1}^{L}\left( \prod_{j\neq k}\frac{u-x_j}{x_k-x_j}\right)\Qt^{-}(x_k)f_{L}\left( \ldots\comma x_k+i\comma \ldots\right)\comma\label{flr3}\\
\bra{\boldsymbol{x}}A_{\epsilon}(u)\ket{\uparrow^{L}}=&\Qt^{+}(u)f_{R}(\boldsymbol{x})\period\label{flr4}
}
By equating \eqref{flr1} with \eqref{flr3} and \eqref{flr2} with \eqref{flr4} and setting $u=x_k$, we can derive the following recursion relations for $f_{L,R}(\boldsymbol{x})$:
\beq{
&f_{L}(\ldots\comma x_k+i\comma \ldots)=\frac{1}{\sqrt{1+\epsilon^2}}f_{L}(\boldsymbol{x})\comma\\
&f_{R}(\ldots\comma x_k-i\comma \ldots)=\frac{1}{\sqrt{1+\epsilon^2}}f_{R}(\boldsymbol{x})\period
}
Solving these recursion relations, $f_{L,R}(\boldsymbol{x})$ can be determined as 
\beq{
f_{L}(\boldsymbol{x})\propto \exp \left( \frac{i}{2}\ln (1+\epsilon^2 ) \sum_{k=1}^{L}x_k\right)\comma\quad f_{R}(\boldsymbol{x})\propto \exp \left( -\frac{i}{2}\ln (1+\epsilon^2 ) \sum_{k=1}^{L}x_k\right)\period\label{fLR}
} 

Let us now convert the summation over the discrete spectrum of $\xhat_k$'s 
 to contour integrals over the continuous variables $x_k$. 
To carry this out, we utilize the following relations:
\beq{
e^{-\pi (x_k-\theta_k)}={\rm Res}_{z=x_{k}}\left[\frac{1}{(z-\theta_k +\frac{i}{2})(z-\theta_k -\frac{i}{2})}\right] \period\label{residue}
}
Note that $x_k$ takes only two values, $\theta_k \pm \frac{i}{2}$ , and \eqref{residue} is either $+i$ or $-i$ depending on which value $x_k$ takes.
%
Then,  by re-expressing the factor $\prod_k e^{-\pi (x_k-\theta_k)}$ in \eqref{sum-mu} using the relations \eqref{residue}, we can rewrite the whole measure as the residue of the following simple function:
\beq{
&\mu (\boldsymbol{x}) \propto {\rm Res}_{\{z_k\}=\{x_k\}}\, \left[  \frac{\prod_{i<j}(z_i-z_j)}{\prod_{l}\Qt^{+}(z_l)\Qt^{-}(z_l)} \right] \comma\label{int-sum}
}
where $\Qt$ is given by $\prod_{k=1}^{L}(u-\theta_k)$ as defined previously in \eqref{Qt}.
The constants of proportionality in \eqref{fLR} and \eqref{int-sum}, which are left undetermined, are functions of the twist parameter $\epsilon$ and the inhomogeneity parameters $\theta_k$'s. These constants are related to the overall normalization of the scalar products and will be fixed by the analysis presented in the 
 Appendix \ref{ap-b}, which compares it with the other known formula for the scalar product.
Taking into account the constants of proportionality in \eqref{fLR} and \eqref{int-sum},  we finally arrive at the following multiple integral formula\fn{In the Appendix \ref{ap-b}, we will give a direct analytical proof of the equivalence between \eqref{asym} and the known determinant formulas.} for the scalar product between an off-shell Bethe-state and an on-shell Bethe state:
 \beq{
 \bra{\uparrow^{L}}\prod_{i=1}^{M}C(v_i)\prod_{j=1}^{M}B(u_j)\ket{\uparrow^{L}} =\frac{\prod_{j<k}(\theta_j-\theta_k )(\theta_j-\theta_k +i) (\theta_j-\theta_k -i)}{(L-2M)!} &\nn\\
 \times \prod_{n=1}^{L}\oint_{\mathcal{C}_n}\frac{dx_n}{2\pi i} \left( \sum_{j=1}^{L}x_j\right)^{L-2M}\prod_{k<l}(x_k-x_l)\prod_{m=1}^{L}\frac{\Qu (x_m)\Qv (x_m)}{\Qt ^{+}(x_m)\Qt ^{-}(x_m)}&\period \label{asym}
 }
In this formula, $\mathcal{C}_n$ denotes  the integration contour which encloses $\theta_n \pm i/2$ counterclockwise. 
Actually the prefactors in front of the integral  are unimportant when computing physical observables,  since they drop out upon normalizing  the Bethe states.
\subsection{Symmetrization and simplification of the multiple integral}
The multiple integral formula \eqref{asym} derived in the last subsection  has one unsatisfactory feature: This expression becomes  singular
as we take the homogeneous limit, $\theta_n\to 0$. There are two sources 
 for the singular behavior. One is that the integration contours  $\mathcal{C}_n$
 get pinched and collide when all the $\theta_n$'s move to the origin. Another source is that
at the same time   the prefactor  $\prod_{j<k}(\theta_j-\theta_k)$  will 
vanish. 
To get around this difficulty, we wish to deform  each integration contour into the 
one, to be denoted by $\mathcal{C}_{\rm all}$,  which encloses all the singularities in the integrand. 
However, if we na\"{i}vely make such deformations,  obviously we will 
 pick up unwanted contributions coming from different integration variables 
 encircling the poles from the same group $\theta_n \pm i/2$. 
We can avoid such contributions by inserting 
 a factor of the form  $\prod_{k<l}(e^{2\pi x_k}-e^{2\pi x_l})$, which vanishes for all the undesired cases.  For the genuine contributions for which this factor 
does not  vanish, we must normalize properly to reproduce the original value 
 of the integral. In this way, with the factor $L!$ coming from the permutation of $x_n$'s, we obtain the following more symmetric expression for the scalar product:
\beq{
&\bra{\uparrow^{L}}\prod_{i=1}^{M}C(v_i)\prod_{j=1}^{M}B(u_j)\ket{\uparrow^{L}} =\frac{\Xi}{L!(L-2M)!}\nn\\
&\hspace{22pt}\times\prod_{n=1}^{L}\oint_{\mathcal{C}_{\rm all}}\frac{dx_n}{2\pi i} \left( \sum_{j=1}^{L}x_j\right)^{L-2M}\!\!\!\!\!\prod_{k<l}(x_k-x_l)(e^{2\pi x_k}-e^{2\pi x_l})\prod_{m=1}^{L}\frac{\Qu (x_m)\Qv (x_m)}{\Qt ^{+}(x_m)\Qt ^{-}(x_m)}\comma\label{sym}
}
where the prefactor $\Xi$ is given by
\beq{
\Xi\equiv \prod_{j<k}\frac{(\theta_j-\theta_k )(\theta_j-\theta_k +i) (\theta_j-\theta_k -i)}{(e^{2\pi \theta_j}-e^{2\pi \theta_k})}\period
}
Note that for this expression the prefactor $\Xi$ is indeed finite in the homogeneous limit. 

Although the expression above is  symmetric in all the variables and 
 hence quite useful, it is of interest  to point out that actually we can 
 integrate out one of the $x_k$'s and obtain a slightly simpler expression 
containing  $L-1$ integration variables. 
To derive it, let us first re-express the factor $\prod_{k<l}(e^{2\pi x_k}-e^{2\pi x_l})$ as a determinant of Vandermonde type:
\beq{
\prod_{k<l}(e^{2\pi x_k}-e^{2\pi x_l}) = \det \left( e^{2\pi (j-1) x_k}\right)_{1\leq j,k\leq L}\period
}
Then, by using the basic definition of the determinant, we can rewrite it 
 into a sum over permutations of the form $\sum_\sigma (-1)^{\sigma} e^{2\pi (\sigma(j)-1) x_{j}}$. Now note that the remaining terms in the integrand is completely antisymmetric with respect to permutations. Hence, all the terms in the above sum  contribute equally and we arrive at the following expression:
\beq{
&\frac{\Xi}{(L-2M)!}\prod_{n=1}^{L}\oint_{\mathcal{C}_{\rm all}}\frac{dx_n}{2\pi i} \left( \sum_{j=1}^{L}x_j\right)^{L-2M}\!\!\!\!\!\prod_{k<l}(x_k-x_l)\prod_{m=1}^{L}\frac{\Qu (x_m)\Qv (x_m)e^{2\pi (m-1) x_m}}{\Qt ^{+}(x_m)\Qt ^{-}(x_m)}\period
} 
Notice that the integral is over meromorphic factors, except for $\exp( 2\pi (m-1) 
x_m)$. However  for $x_1$ this factor is absent. Hence we can easily integrate 
 out this variable by closing its contour at infinity. At infinity 
 all the factors become 
  power functions  and the only non-vanishing integral to be performed is $\oint dx_1/(2\pi i x_1) =1$.  After this procedure, we may convert the 
 factor  $e^{2\pi (m-1) x_m}$ back to the determinant and further to 
 the original expression  $\prod_{k<l}(e^{2\pi x_k}-e^{2\pi x_l})$. 
In this way  we obtain the following simple formula with $L-1$ integration variables:\fn{For simplicity, we have renamed $x_2,\ldots ,x_L$ as $x_1,\ldots ,x_{L-1}$.}
\beq{
&\bra{\uparrow^{L}}\prod_{i=1}^{M}C(v_i)\prod_{j=1}^{M}B(u_j)\ket{\uparrow^{L}} =\frac{\Xi}{(L-1)!(L-2M)!}\nn\\
 &\hspace{33pt}\times\prod_{n=1}^{L-1}\oint_{\mathcal{C}_{\rm all}}\frac{dx_n}{2\pi i} \prod_{k<l}(x_k-x_l)(e^{2\pi x_k}-e^{2\pi x_l})\prod_{m=1}^{L-1}\frac{\Qu (x_m)\Qv (x_m)e^{2\pi x_m}}{\Qt ^{+}(x_m)\Qt ^{-}(x_m)}\period\label{L-1}
} 
Note that  the factor, $(\sum_j x_j)^{L-2M}$, which was present in the previous expressions, disappeared upon integration over $x_1$. Therefore \eqref{L-1} is structurally similar to the eigenvalue integral of a matrix model. Namely,  $Q$-functions correspond to a potential term for the eigenvalues and $\prod_{k<l}(x_k-x_l)(e^{2\pi x_k}-e^{2\pi x_l})$ can be interpreted as a modified Vandermonde factor. It is intriguing to note that this modified Vandermonde factor 
is a hybrid of  the ordinary Vandermonde factor for the Hermitian matrix model, $\prod_{k<l}(x_k-x_l)^2$, and  the generalized Vandermonde factor for the unitary matrix model and the Chern-Simons matrix model \cite{AKMV}, which is essentially given by $\prod_{k<l}(e^{2\pi x_k}-e^{2\pi x_l})^2$. This  resemblance to a matrix model strongly suggests that the semi-classical limit for the scalar product can be analyzed by applying the 
method of  large $N$ expansion familiar for matrix models. 
It turns out, however, that the separated variables we deal with here 
and the eigenvalues of a matrix model behave somewhat differently and  
one must be very careful in adapting such a method. This subject will be 
 discussed further in the final section. 
\section{Discussions\label{sec-5}}
In this paper, applying the method of separation of variables \`a la Sklyanin, 
we have derived a multiple contour integral representation for  the scalar products
 between an on-shell and an off-shell 
 Bethe states of the XXX spin $1/2$ chain with periodic boundary condition. 
Although a compact determinant expression for such a scalar product 
 was discovered more than two decades ago \cite{Slavnov} and various variants 
have been discussed since then, our novel  integral formula should be useful for 
 a number of purposes. One such area is the study of the semi-classical 
 behavior of the scalar products. In particular, this is quite 
 important for the comparison of the three point functions 
 in super Yang-Mills theory and the corresponding dual string theory. 
Already  results  have  been obtained  in this regard \cite{EGSV3, Kostov, Kostov2}
 starting from the determinantal form of the scalar product. However, the 
procedure through which these results are obtained is ingenious but not
 quite systematic. The purpose of our present work is largely to improve on 
 this situation. 

There are indeed apparent advantages for our formula for such a purpose. 
Firstly, almost by definition, the SoV method we employed guarantees that 
 the factorization of the dependence on the basic  variables occur. 
Such a factorization is not realized in the determinant formula and indeed 
 it took some non-trivial steps to obtain such a structure in \cite{Kostov}.
A further merit of the inherent factorization is  that  the factorized dependence
 on the inhomogeneous parameters $\theta_k$  may  be quite useful in 
applying the so-called $\Theta$-morphism operation \cite{GV,EGSV4}, 
 which could  be a key for understanding the loop effects. (See also \cite{Serban1,Serban2}.) 

Another advantage of our formula is that, as was shown in section \ref{sec-4}, 
it can be put into a form reminiscent 
 of the integral over the eigenvalues of a matrix model. 
An intriguing fact  is that the measure factor of our integral formula 
looks like a hybrid of that of a hermitian  matrix model and 
a unitary matrix model. 
The semi-classical  limit of interest corresponds to the large size limit of the matrix and 
various techniques developed in the past may be utilized. 
However, preliminary investigation indicates that there are certain 
 important differences between our integral and the matrix integral. 
For instance, in contrast to the usual matrix model eigenvalues, 
  the separated variables $x_k$'s do not necessarily condense. It appears 
 that  careful analysis for the various regions of $x_k$'s is required in order 
 to extract the semi-classical contributions systematically.  Such an analysis 
 will be presented elsewhere \cite{KKN}. 

Apart from the semi-classical limit, our formalism should be useful for the general development of solvable models.  An immediate 
 application may be  to the SU(3) spin chain, for which the SoV method is 
 available \cite{sl(3),sl(N)}. It is of interest to see what modification from the SU(2) case 
 would be  needed to obtain a nice integral formula for the scalar
 product for such a system. 

Before ending this article,  we should perhaps emphasize, alongside with Sklyanin, that the method of SoV, or functional Bethe ansatz as it was originally 
 called,  is in a sense the most fundamental and general 
of all the Bethe ansatz methods and hence should be applicable to 
systems to which the other methods are not easily applicable. 
In this sense, we hope that our analysis will prove to be useful for future development of the SoV method. 
\par\bigskip\noindent
{\large\bf Acknowledgment}\par\smallskip\noindent
We thank P.~Vieira for useful discussions.
The research of  Y.K. is supported in part by the 
 Grant-in-Aid for Scientific Research (B) 
No.~20340048 and (B) No.~25287049, while that of S.K. is supported in part 
 by JSPS Research Fellowship for Young Scientists,   from the Japan 
 Ministry of Education, Culture, Sports,  Science and Technology. 
\appendix
\section{Proof of theorem 1\label{ap-a}}
In this appendix, we shall provide  a proof of the theorem \hyperref[theorem]{1}, which gives the action of the normal-ordered 
 operators $\text{:}A_\ep(\hat{x}_k)\text{:}_L$  and $\text{:}D_\ep(\hat{x}_k)\text{:}_L$
 on the SoV bra $\langle \ldots, x_k, \ldots |$.  Essentially, what follows is 
 a pedagogical adaptation of the argument given in \cite{Sklyanin2}. 

The proof is by mathematical induction in  the number of sites $L$. 
Begin with the  $L=1$ case.  The operators $A_\ep(u),B_\ep(u),C_\ep(u),D_\ep(u)$ are given by\footnote{We have dropped the subscript 1 for $S^k$ for simplicity. Also subscripts $\ep$ for $A(u),\ldots ,D(u)$ are suppressed.}
\begin{align}
A(u) &= u-\theta_1 + iS^z + i\ep S^+\comma  &B(u) &= \ep (u-\theta_1 -iS^z) + iS^-\comma \\
C(u) &= -\ep ( u-\theta_1 + iS^z) + iS^+\comma &D(u)& = u-\theta_1 -iS^z -i\ep S^-\period
\end{align}
By solving  $B(\xhat_1)=0$ for $\xhat_1$ and substituting it into 
 $A(u)$ and $D(u)$, we get 
\begin{align}
\xhat_1 &= \theta_1 + iS^z -i\ep^{-1}S^- =\matrixii{\theta_1+{i \over 2}}{0}{-i \ep^{-1}}{\theta_1 -{i \over 2}} \comma \\
\text{:}A(\xhat_1)\text{:}_L  &= 2i S^z -i\ep^{-1}S^- + i\ep S^+ =i \matrixii{1}{\ep}{-\ep^{-1}}{-1} 
\comma \\
\text{:}D(\xhat_1)\text{:}_L  &=-i(\ep + \ep^{-1}) S^- =-i (\ep + \ep^{-1})
\matrixii{0}{0}{1}{0} \period
\end{align}
Since $\xhat_1$ is lower triangular, its eigenvalues are read off as  $\theta_1 \pm {i \over 2}$ and the corresponding normalized eigenbras $\bra{\pm}$ are given by 
$\bra{+} = (1,0), \bra{-}= (1,\ep)/\sqrt{1+\ep^2} $. Then, we can compute the action of $\text{:}A(\xhat_1)\text{:}_L$ 
and $\text{:}D(\xhat_1)\text{:}_L $ explicitly and get
\begin{align}
\bra{+} \text{:}A(\xhat_1)\text{:}_L &= i \sqrt{1+\ep^2}\bra{-} \comma \quad 
\bra{-} \text{:}A(\xhat_1)\text{:}_L =0 \comma \label{L1A}\\
\bra{+} \text{:}D(\xhat_1)\text{:}_L &= 0 \comma \qquad \hspace{1.5cm}
\bra{-}  \text{:}D(\xhat_1)\text{:}_L = -i \sqrt{1+\ep^2}   \bra{+} \period \label{L1D}
\end{align}
This is precisely what the theorem says for $L=1$. 

Next, assume that the formulas hold for up to $L=N$ and consider $L=N+1$ 
 case. The monodromy matrix for $L=N+1$ is given by 
\begin{align}
\Omega_{N+1} &= K L_1 \cdots L_N L_{N+1} \comma 
\end{align}
where $K$ is the twisting matrix given  by $K =\matrixii{1}{\epsilon }{-\epsilon }{1} $. Now in order to split this 
into the monodromy matrix at the $N$th level  and the subsequent 
action of the Lax operator at the $(N+1)$th step, we should introduce  in the final Lax operator a  twisting 
 matrix of a similar form, which we denote as 
\begin{align}
\Ktil &\equiv \matrixii{1}{\eta}{-\eta}{1} \comma 
\end{align}
with  $\eta$ being  an arbitrary parameter, just like $\ep$. 
Then we can write 
$\Omega_{N+1} = \tilde{\Omega}_N L'_{N+1} $, 
where
\begin{align}
\tilde{\Omega}_N &= (K\Ktil^{-1}) \Ltil_1 \Ltil_2 \cdots \Ltil_N \comma \qquad K \Ktil^{-1} =\frac{1}{1+\eta^2} \matrixii{1+\ep \eta}{\ep -\eta}{-(\ep -\eta)}{1+\ep \eta} \comma \\
\Ltil_i &= \Ktil L_i \Ktil^{-1} \comma \qquad 
L'_{N+1} = \Ktil L_{N+1}\period
\end{align}
Since the conjugation by $\Ktil$ does not affect the structure of the algebra, 
 we may regard $\tilde{\Omega}_N$ as the monodromy matrix for $L=N$ 
for which the theorem holds with the factor $\sqrt{1+\ep^2}$ in \eqref{bra-A} and \eqref{bra-D} replaced with $\sqrt{(1+\ep^2)/(1+\eta^2)}$. We now write the matrix elements of $\tilde{\Omega}_N$
 and $L'_{N+1}$ as 
\begin{align}
\tilde{\Omega}_N (u) &= \matrixii{\Atil_N}{\Btil_N}{\Ctil_N}{\Dtil_N} 
\comma \qquad L'_{N+1} = \matrixii{a_{N+1}}{b_{N+1}}{c_{N+1}}{d_{N+1}} \comma
\end{align}
and compute $\Omega_{N+1} = \matrixii{A_{N+1}}{B_{N+1}}{C_{N+1}}{D_{N+1}} $. Then $B_{N+1} $ operator is given by 
\begin{align}
B_{N+1}(u) &= \Atil_N(u) b_{N+1}(u) + \Btil_N(u) d_{N+1}(u) 
\period
\end{align}
Let us now write the SoV basis of bras for $L=N+1$ 
as $\bra{x_1, \ldots, x_N; y}$. By the hypothesis of the induction, 
 $\Btil_N(u)$ is diagonal in this basis and also $b_{N+1}$, which acts only 
 on the $(L+1)$th  site,   is diagonal. Explicitly, we have
\begin{align}
\bra{x_1, \ldots, x_N; y} \Btil_N(u) &= \frac{\ep -\eta}{1+\eta^2} \prod_{i=1}^N
(u-x_i) \bra{x_1, \ldots, x_N; y} \comma \\
\bra{x_1, \ldots, x_N; y} b_{N+1}(u) &= \eta (u-y) \bra{x_1, \ldots, x_N; y}
\period
\end{align}
We may now compute the action of $\Btil_{N+1}(u)$ at $u=\xhat_k$ and 
$u=\hat{y}$, where $\hat{y}$ is the root of $b_{N+1}(\hat{y}) =0$ given by 
$\hat{y} = \theta_{N+1} +iS^3_{N+1} -i\eta^{-1} S^-_{N+1}$. 
Since $\Btil_N(\xhat_k)$ and $b_{N+1}(\hat{y})$ vanishes on this state, 
we get
\begin{align}
\bra{x_1, \ldots, x_N; y} B_{N+1}(\xhat_k) &= \bra{x_1, \ldots, x_N; y}\Atil_N(\xhat_k) b_{N+1}(\xhat_k) \comma \\
\bra{x_1, \ldots, x_N; y}B_{N+1}(\hat{y}) &= \bra{x_1, \ldots, x_N; y}\Btil_N(\hat{y}) 
 d_{N+1} (\hat{y}) \period
\end{align}
The RHS can be easily computed since $\Atil_N(\xhat_k)$ shifts 
  $x_k$ by $-i$,  while $d_{N+1}(\hat{y})$ shifts $y$ by $+i$, with 
 certain known factors multiplied. In this way, we obtain the formulas 
\begin{align}
&\bra{x_1, \ldots, x_N; y} B_{N+1}(\xhat_k) = \eta \sqrt{\frac{1+\ep^2}{1+\eta^2}}(x_k-y) 
Q^+_{\boldsymbol{\theta}}(x_k) \langle \ldots, x_k-i , \ldots;y |
\comma 
\\
&\bra{x_1, \ldots, x_N; y}B_{N+1}(\hat{y}) =\frac{\ep -\eta}{1+\eta^2}\sqrt{\frac{1+\ep^2}{1+\eta^2}} 
(y-\theta_{N+1} -i/2) \prod_{k=1}^N (y-x_k) 
 \langle \ldots, x_k , \ldots;y+i | \period
\end{align}

Having understood the action of $B_{N+1}$ at $u=\xhat_k, \yhat$ on the SoV basis, we now wish to deduce the spectrum of $B_{N+1}(u)$ using this information. 
Let $\ket{\Phi}$ be the state which diagonalizes $B_{N+1}(u)$. Then 
by taking the inner product with the above two equations, we obtain
\begin{align}
&\beta (x_k) \Phi (x_1, \ldots, x_n; y) = \eta \sqrt{\frac{1+\ep^2}{1+\eta^2}}(x_k-\eta) 
Q^+_{\boldsymbol{\theta}}(x_k) \Phi(\ldots, x_k-i , \ldots;y )
\comma 
\\
&\beta (y) \Phi(x_1, \ldots, x_N; y)=\frac{\ep -\eta}{1+\eta^2}\sqrt{\frac{1+\ep^2}{1+\eta^2}} 
(y-\theta_{N+1} -i/2) \prod_{k=1}^N (y-x_k) 
 \Phi(\ldots, x_k , \ldots;y+i ) \comma 
\end{align}
where $\Phi(x_1,\ldots, x_N ;y ) \equiv \langle x_1, \ldots, x_N;y \ket{\Phi}$
 and we have denoted the eigenvalue of $B_{N+1}(u)$ by $\beta (u)$. 
Now to simplify the analysis of the spectrum, it is convenient to 
extract a factor $\rho(x_1, \ldots, x_N; y)$ from $\Phi(x_1, \ldots, x_N; y)$
 in the manner 
\begin{align}
\Phi(x_1, \ldots, x_N; y) &= \rho(x_1, \ldots, x_N; y) \Psi(x_1, \ldots, x_N; y)
\end{align}
where $\rho(x_1, \ldots, x_N; y)$ satisfies the  first order difference 
equations 
\begin{align}
 \rho(x_1, \ldots, x_N; y) &= \eta \sqrt{\frac{1+\ep^2}{1+\eta^2}} (x_k-\eta)  \rho( \ldots, x_k-i, \ldots; y)
\comma \\
 \rho(x_1, \ldots, x_N; y) &=\frac{\ep -\eta}{1+\eta^2}\sqrt{\frac{1+\ep^2}{1+\eta^2}}
\prod_{i=1}^N(y-x_k) \rho( \ldots, x_k, \ldots; y+i) \period
\end{align}
One can easily verify that the solution to these equations is unique\footnote{
The uniqueness is guaranteed by the finiteness of the spectrum of $\xhat_k$
 and $\yhat$. One can construct the solution $\rho$ by starting from 
 the end  of the spectrum.}
 up to  an overall constant.  Now with such a factor removed, 
the reduced wave function $\Psi$ satisfies the equations 
\begin{align}
\beta (x_k) \Psi(x_1, \ldots, x_N; y) &= Q^+_{\boldsymbol{\theta}}(x_k) 
 \Psi(\ldots, x_k-i, \ldots; y) \comma \\
\beta (y) \Psi(x_1, \ldots, x_N; y) &= (y-\theta_{N+1} -i/2) 
 \Psi(\ldots, x_k, \ldots; y+i) \period
\end{align}
It turns out that we can drastically  simplify these equations
 by assuming the factorized 
 form\footnote{This does not miss any solution since the  solution is unique.}
 for $\Psi$, namely
\begin{align}
\Psi(x_1, \ldots, x_N;y) &= \chi(y) \prod_{k=1}^N \xi_k(x_k) \period
\end{align}
The equations for $\Psi$ then get reduced to the following equations for
each  factor
\begin{align}
\beta (x) \xi_k(x) &= Q^+_{\boldsymbol{\theta}}(x) \xi_k(x-i) 
\comma \qquad\hspace{2cm} x \in \left\{ \theta_k -{i \over 2}, \theta_k + {i \over 2}\right\} \label{xieq}
\comma \\
\beta (x) \chi(x) &= (y-\theta_{N+1} -i/2) \chi(x+i) \comma \qquad 
x \in \left\{ \theta_{N+1}-{i \over 2}, \theta_{N+1} + {i \over 2}\right\}
\period \label{chieq}
\end{align}
Note that we have used the induction hypothesis  that the spectrum of each $x_k$ is two-valued as above.

 The rest of the analysis is elementary. First 
 consider the equation (\ref{xieq}) and set  $x=\theta_k -{i \over 2}$. Then due to the presence of the factor $Q^+_{\boldsymbol{\theta}}(x)$ the RHS vanishes and hence we must have  $\beta (\theta_k-{i\over 2}) \xi_k (\theta_k-{i\over 2})  =0$. 
If $\xi_k (\theta_k-{i\over 2})\ne 0$,  then $\beta (\theta_k-{i\over 2})$
 must vanish and  $\theta_k-{i \over 2}$ is in the spectrum. 
On the other hand suppose  $\xi_k (\theta_k-{i\over 2})=0$. Then 
  $\xi_k (\theta_k+{i\over 2})$ cannot vanish since otherwise the whole 
 wave function vanishes. Now set $x=\theta_k+{i\over 2}$ in (\ref{xieq}). 
 Then the RHS vanishes and so must the LHS, \ie  $\beta (\theta_k+{i\over 2}) \xi_k (\theta_k+{i\over 2})  =0$. This leads to  $\beta (\theta_k+{i\over 2})=0$ and hence 
 $x=\theta_k+{i\over 2}$ is in the spectrum. Similar arguments for 
(\ref{chieq}) tells us that $\theta_{N+1} \pm {i \over 2}$ are in the 
 spectrum. Thus, for  $L=N+1$, we continue to have the same set of spectrum 
 as stated in the theorem. 

From this analysis we learn that the finite discrete 
 nature of the spectrum is due to two reasons. One is that the operators 
 $\text{:}A(\hat{x}_k)\text{:}_L$ and $\text{:}D(\hat{x}_k)\text{:}_L$ are essentially exponentials  of the momentum 
 operator and hence they induce a finite shift in $x_k$. The second ingredient 
 is the presence of the prefactor $Q^+_{\boldsymbol{\theta}}(x)$. 
 Since it vanishes at finite discrete values of $x$, the shifting must end after a finite 
 number of steps, in the present case just one. 

What remains  is the determination of the constant of 
 proportionality in the action of the operators $\text{:}A(\hat{x}_k)\text{:}_L$ and $\text{:}D(\hat{x}_k)\text{:}_L$. 
As there are only a finite number of states, such a constant can be adjusted 
 rather freely by the change of the normalization of states. Nonetheless, 
 there is a certain constraint coming from the following non-linear 
 relations: 

\begin{align}
\text{:}A_{N+1}(\xhat_k)\text{:}_L \text{:}D_{N+1}(\xhat_k)\text{:}_L
&= (1+\ep^2) \prod_{l=1}^{N+1} (\xhat_k -\theta_l + i/2)(\xhat_k -\theta_l - 3i/2)  \comma \label{ADeq}\\
\text{:}D_{N+1}(\xhat_k)\text{:}_L \text{:}A_{N+1}(\xhat_k)\text{:}_L
&= (1+\ep^2) \prod_{l=1}^{N+1} (\xhat_k -\theta_l - i/2)(\xhat_k -\theta_l + 3i/2) \period \label{DAeq}
\end{align}
These relations can be obtained in the following way. From the 
 commutation relations between $\text{:}A_{N+1}(\xhat_k)\text{:}_L$, 
$ \text{:}D_{N+1}(\xhat_k)\text{:}_L$ and $\xhat_k$, one can show 
\begin{align}
\text{:}A_{N+1}(\xhat_k):_L :D_{N+1}(\xhat_k)\text{:}_L &= \mbox{$\det_q$} \Omega_{N+1}(\xhat_k
-i / 2) \comma\\
\text{:}D_{N+1}(\xhat_k)\text{:}_L \text{:}A_{N+1}(\xhat_k)\text{:}_L 
&= \mbox{$\det_q$} \Omega_{N+1} ( \xhat_k +i/2)\comma
\end{align}
where $\det_q \Omega_{N+1}(u)$ is the so-called 
quantum determinant\footnote{For a detailed account of the quantum determinant, we refer the reader to \cite{book} and \cite{Sklyanin2}.}
, which is a central element of the Yang-Baxter exchange 
 algebra. Then by using the co-multiplication rule,  $\det_q (AB) = \det_q A\, 
 \det_q B$, one can explicitly compute the RHS and obtain the 
 relations (\ref{ADeq}) and (\ref{DAeq}).  The constant of proportionality
 chosen in the theorem is compatible with these relations and also to 
the explicit equations for $L=1$ case shown in (\ref{L1A}) and 
 (\ref{L1D}) obtained for unit-normalized states. 
This completes the proof of the theorem. 
\section{Relation to Izergin's determinant formula\label{ap-b}}
In this appendix, we give a direct proof that a slight generalization of our new integral expression  is equivalent to the Izergin's 
determinant formula \cite{Izergin} for the domain wall partition function (DWPF)
 which appears in the six-vertex model. 
From this DWPF, the original scalar product of our interest can  be obtained by 
sending an appropriate subset of rapidities to infinity 
as well as requiring  half of the remainder to be on-shell.  

We begin with the domain wall partition function, which is defined 
 as follows:
\begin{align}
Z_L({\boldsymbol w}|\boldsymbol{ \theta })  \equiv \langle \downarrow^L | \prod_{i=1}^L B(w_i) | \uparrow^L \rangle \label{eq:DWPF} \period
\end{align}
Note that the number of $B$ operators is equal to the number of sites $L$ and 
the rapidities ${\boldsymbol w}$ are not restricted to an on-shell configuration.
In \cite{Izergin}, Izergin gave a determinant expression for this quantity, which reads 
\begin{align}
Z_L(\boldsymbol{ w}|\boldsymbol{ \theta } )=  \frac{\prod_{j,k=1}^L (w_j-\theta_k +\frac{i}{2})(w_j-\theta_k -\frac{i}{2})}{\prod_{l <m } (w_l -w_m )(\theta_m -\theta_l) } \det \left( \frac{i}{(w_j-\theta_k+\frac{i}{2})(w_j-\theta_k -\frac{i}{2})} \right)_{1\leq j,k\leq L} \period
\label{eq:Izergin}
\end{align}
In what follows, we shall show that this is equal to the multiple integral formula of the form 
\begin{align}
\langle \downarrow^L | \prod_{i=1}^L B(w_i) | \uparrow^L \rangle&= i^L \prod_{j<k} (\theta_j -\theta_k)(\theta_j -\theta_k+i)(\theta_j -\theta_k-i) \nonumber \\
&\times \prod_{n=1}^{L} \oint_{\mathcal{C}_n} \frac{dx_n}{2\pi i} \prod_{k<l} (x_k-x_l) \prod_{m=1}^L \frac{Q_{\boldsymbol{ w}}(x_m)}{Q_{\boldsymbol{\theta}}^{+}(x_m)Q_{\boldsymbol{\theta}}^{-}(x_m)}  
\period
\label{eq:MCI}
\end{align}

First, we shall transform the Izergin's formula to a form more convenient 
 for comparison with the integral expression.  By a simple decomposition, 
 the determinant in (\ref{eq:Izergin}) can be rewritten as a determinant of 
the difference of two matrices:
\begin{align}
\det \left(  \frac{i}{(w_j-\theta_k+\frac{i}{2})(w_j-\theta_k -\frac{i}{2})} \right)_{1\leq j,k\leq L} &=\det (M_{jk}^- - M_{jk}^+)_{1\leq j,k\leq L}
\comma \label{eq:det} \\
M_{jk}^{\pm} &={1\over w_j-\theta_k\pm i/2 } \period
\end{align}
Then from the definition of the determinant, we can expand the RHS of (\ref{eq:det}) as 
\begin{align}
\det (M_{jk}^- - M_{jk}^+)_{1\leq j,k\leq L}&=\sum_{\sigma \in P_L}(-1)^{\sigma}(M_{1\sigma(1)}^- - M_{1\sigma(1)}^+)\cdots (M_{L\sigma(L)}^- - M_{L\sigma(L)}^+) \nonumber \\
&=\sum_{\epsilon_i =\pm} (-1)^{n_+} \sum_{\sigma \in P_L}(-1)^{\sigma} M_{1\sigma(1)}^{\epsilon_1} \cdots M_{L\sigma(L)}^{\epsilon_L}
\comma 
\label{eq:sum}
\end{align}
where $n_+$ is the number of $+$'s  in the set $\{ \epsilon_i \}$ and the sign $(-1)^{n_+}$ is produced  upon  expanding the product. 
Now by using the definition of determinant again to re-express 
 each summand  back as a determinant,  we obtain 
\begin{align}
\det (M_{jk}^- - M_{jk}^+)_{1\leq j,k\leq L}=\sum_{\epsilon_i =\pm} (-1)^{n_+} \det (M_{jk}^{\epsilon_j})_{1\leq j,k\leq L} \period
\label{eq:sumdet}
\end{align}
At this point,  one can apply the Cauchy's determinant identity,
\begin{align}
\det \left( \frac{1}{x_j-y_k} \right)_{1\leq j,k\leq L} = \frac{\prod_{1\leq j<k\leq L}(x_j-x_k)(y_k-y_j)}{\prod_{l,m=1}^L(x_l-y_m)}\comma
\label{eq:Cauchy}
\end{align}
to each term $ \det (M_{jk}^{\epsilon_j})_{1\leq j,k\leq L}=\det ( (w_j-(\theta_k-\epsilon_j i/2))^{-1} )_{1\leq j,k\leq L}$.  
Putting altogether the determinant (\ref{eq:det}) in the Izergin's formula 
can be expressed  as 
\begin{align}
\det \left(  \frac{i}{(w_j-\theta_k+\frac{i}{2})(w_j-\theta_k -\frac{i}{2})} \right)_{1\leq j,k\leq L}=\sum_{\ep_i=\pm}(-1)^{n_+}\frac{\prod_{1\leq j<k\leq L}(w_j-w_k)(\theta_k-\theta_j-i(\epsilon_k-\epsilon_j)/2)}{\prod_{l,m=1}^L(w_l-\theta_m+i\epsilon_m /2)}
\period
\end{align}
Substituting it into (\ref{eq:Izergin}),  Izergin's formula is finally transformed into 
the expression
\begin{align}
Z_L(\boldsymbol{ w}|\boldsymbol{ \theta } )= \sum_{\epsilon_i =\pm} (-1)^{n_+} \prod_{j,k=1}^L(w_j-(\theta_k+i\epsilon_k /2))\prod_{1\leq l<m\leq L}\frac{(\theta_l-\theta_m-i(\epsilon_l-\epsilon_m)/2)}{\theta_l -\theta_m}
\comma 
\label{eq:sum_p}
\end{align}
which is no longer of a determinant form. 

Now we are ready to prove its  equivalence to the 
 multiple integral (\ref{eq:MCI}). This is done essentially by 
explicitly performing the contour integrals using the residue formula. 
By picking up the contributions from the zeros of the functions $Q^\pm_{\boldsymbol{\theta}}(x_m)$ in the denominator,  the integral is evaluated as 
\begin{align}
&\prod_{r<s} (\theta_r -\theta_s)(\theta_r -\theta_s+i)(\theta_r -\theta_s-i) \nonumber \\
&\times \sum_{\epsilon_i =\pm} (-1)^{n_+} \prod_{1\leq l<m\leq L}\frac{(\theta_l-\theta_m+i(\epsilon_l-\epsilon_m)/2)}{(\theta_l -\theta_m)^2(\theta_l -\theta_m+\epsilon_l)(\theta_l -\theta_m-\epsilon_m)} \prod_{j,k=1}^L(w_j-(\theta_k+i\epsilon_k /2))
\period\label{eq:sum_p2}
\end{align}
Now note the following relation, which can be checked for 
every pair $(\epsilon_l, \epsilon_m)$, with $\epsilon_l =\pm 1$:
\begin{align}
\frac{(\theta_l-\theta_m+i)(\theta_l-\theta_m-i)}{(\theta_l-\theta_m+i\epsilon_l)(\theta_l-\theta_m-i\epsilon_m)}=\frac{(\theta_l-\theta_m-i(\epsilon_l-\epsilon_m)/2)}{(\theta_l-\theta_m+i(\epsilon_l-\epsilon_m)/2)} \period
\end{align}
Using this formula,  the expression (\ref{eq:sum_p2}) can be  simplified into 
\begin{align}
 \sum_{\epsilon_i =\pm} (-1)^{n_+} \prod_{j,k=1}^L(w_j-(\theta_k+i\epsilon_k /2))\prod_{1\leq l<m\leq L}\frac{(\theta_l-\theta_m-i(\epsilon_l-\epsilon_m)/2)}{\theta_l -\theta_m} \period
\end{align}
This is exactly the same as (\ref{eq:sum_p}), proving the assertion. 

As already stated, the original scalar product of our interest 
 can be obtained from this  domain wall partition function through certain 
 manipulations. First, by sending $L-n$ of the $L$ rapidities to 
 infinity, thereby decoupling them, one obtains the 
 partial domain wall partition function with $n$ rapidities 
 $\boldsymbol{z}$ \eqref{eq:third} in the following way:
\begin{align} Z^{\rm{pDWPF}}(\boldsymbol{z}|\boldsymbol{ \theta } )=\frac{1}{(L-n)!} \lim_{\{ w_1,\ldots , w_{L-n} \}\rightarrow \infty} \left( \frac{Z_L(\boldsymbol{z} \cup \{ w_1,\ldots ,w_{L-n} \} |\boldsymbol{ \theta} )}{ iw_1^{L-1} \cdots iw_{L-n}^{L-1}} \right) 
\period
\label{eq:pDWPF}
\end{align}
If we now set $n=2M$ and $\boldsymbol{z}=\boldsymbol{u}\cup \boldsymbol{v}$,  where either $\boldsymbol{u}$ or $\boldsymbol{v}$ are on-shell, we recover the original scalar product $\langle \uparrow^L | \prod_{i=1}^M C(v_i) \prod_{j=1}^M B(u_j) |\uparrow^L \rangle$.  On the other hand, if we apply the same manipulations to the integral formula \eqref{eq:Izergin}, we obtain the multiple integral formula for the scalar product \eqref{asym}. This proves the equivalence of our formula and the determinant formula derived by Foda and Wheeler \cite{FW2}.

\end{document}